\documentclass[journal]{IEEEtran}
\usepackage[cmex10]{amsmath}
\usepackage{amsfonts}
\usepackage{amssymb}
\usepackage{bm}
\usepackage{algorithmic}
\usepackage{algorithm}
\usepackage{array}

\usepackage{textcomp}
\usepackage{stfloats}
\usepackage{url}
\usepackage{tikz}
\usepackage{verbatim}
\usepackage{graphicx}
\usepackage{cite}

\usepackage[subtle,tracking=normal]{savetrees}
\usepackage{soul}
\usepackage[dvipsnames]{xcolor}
\sethlcolor{CornflowerBlue}

\usepackage{multirow}

\usepackage[tight,footnotesize]{subfigure}
\usepackage{subfiles}
\usepackage{pdfpages}

\newtheorem{definition}{Definition}
\newtheorem{proposition}{Proposition}
\newtheorem{remark}{Remark}

\newcommand{\rev}[1]{\textcolor{black}{#1}}
\begin{document}

\title{Codeword-Segmentation Rate-Splitting Multiple Access and Evaluation under Suboptimal Decoding}

\author{
    \IEEEauthorblockN{
        Sibo~Zhang,~\IEEEmembership{ Member,~IEEE}, Bruno~Clerckx,~\IEEEmembership{Fellow,~IEEE}}, David Vargas% <-this % stops a space
    \thanks{This work was supported in part by the UK Engineering and Physical Sciences Research Council, Industrial Case award number 210163. (Corresponding author: Bruno Clerckx.)}
    \thanks{S. Zhang is with the Department of Electrical and Electronic Engineering at Imperial College London, London SW7 2AZ, U.K. (e-mail: siboz0477@gmail.com).}
    \thanks{B. Clerckx is with the Department of Electrical and Electronic Engineering at Imperial College London, London SW7 2AZ, U.K., and also with Kyung Hee University, Seoul, South Korea  (e-mail: b.clerckx@imperial.ac.uk).}
    \thanks{D. Vargas is with BBC Research and Development, The Lighthouse, White City Place, 201 Wood Lane, London, W12 7TQ, U.K. (e-mail: david.vargas@bbc.co.uk).} 
    }% <-this % stops a space



\maketitle

\begin{abstract}
Rate-Splitting Multiple Access (RSMA) has been recognized as a promising multiple access technique. We propose a novel architecture for downlink RSMA, namely Codeword-Segmentation RSMA (CS-RSMA). Unlike conventional RSMA, which splits users' messages into common and private parts before encoding, CS-RSMA encodes the users' messages directly, segments the codewords into common and private parts, and transmits the codeword segments over common and private streams. In addition to the principle of CS-RSMA, a novel performance analysis framework is proposed. \rev{This framework leverages a recent discovery in mismatched decoding under finite-alphabet input and interference and can more accurately capture the receiver's complexity limits.} Precoder optimization under finite alphabets and suboptimal decoders for conventional RSMA and CS-RSMA to maximize the Sum-Rate (SR) and the Max-Min Fairness (MMF) is also addressed. The numerical results reveal the theoretical performance of conventional RSMA and CS-RSMA. We observe that CS-RSMA yields better performance than conventional RSMA in SR and similar performance in MMF. Furthermore, a physical-layer implementation of CS-RSMA is proposed and evaluated through link-level simulations. Aside from performance benefits, we demonstrate that CS-RSMA brings significant benefits to the encoding/decoding, control signaling, and retransmission process compared to conventional RSMA.
\end{abstract}


\section{Introduction}

Emerging wireless networks, such as 6G, face increasing demands for high spectral efficiency and user fairness in massive connectivity. Such demands require the development of advanced multiple access techniques \cite{Clerckx_Proceedings_IEEE, Jorswieck_survey}. Rate-Splitting Multiple Access (RSMA) has arisen as a promising paradigm due to its numerous advantages in terms of efficiency, reliability, robustness, low latency, flexibility and universality \cite{Clerckx_JSAC}. The key idea of downlink RSMA involves combining part of users' messages into common streams, which are transmitted in a multicast fashion, whereas the remaining messages are transmitted through private streams in a unicast fashion. The existence of a mixture of common and private streams provides flexibility and improved interference management capability compared to conventional techniques, such as Orthogonal Multiple Access (OMA), Space-Division Multiple Access (SDMA) and Non-Orthogonal Multiple Access (NOMA), as demonstrated theoretically and numerically \cite{Clerckx_JSAC}, and experimentally \cite{Lyu_TWC}.

In the majority of the existing work on RSMA, Successive Interference Cancellation (SIC) is assumed to be implemented at the receivers to separate the common and private streams. More recently, \cite{Sibo_TCOM} considers RSMA under finite alphabets and proposed non-SIC receivers for RSMA and the corresponding transmitter design, also called SIC-free RSMA.\footnote{Previous studies use the terms ``non-SIC" and ``SIC-free" interchangeably, as they convey the same meaning. \rev{In this work, we adopt the term "SIC-free" for consistency.}} Although most prior works analyze RSMA under Gaussian-distributed inputs, the analysis and evaluation of RSMA under finite alphabets are motivated by two factors: (1) most practical systems operate with finite alphabets due to the ease on hardware requirement and encoding/decoding; (2) finite-alphabet interference is less detrimental than Gaussian-distributed interference, it is therefore possible to preserve the benefit of having a mixture of common and private streams without removing the common streams. In \cite{Sibo_TCOM}, \rev{SIC-free} RSMA has been shown to lead to savings in receiver complexity and latency, but only with a minor sacrifice in both theoretical performance, measured by constellation-constrained mutual information, and practical performance, measured by Link-Level Simulations (LLS). \cite{Sibo_SPAWC, Sibo_TWC} further propose precoder optimization for SIC-based and \rev{SIC-free} RSMA under finite alphabets. The results verified that RSMA's advantage is preserved for finite alphabets, and that SIC-free RSMA incurs only a minor loss compared to RSMA with SIC.

SDMA is currently the predominant spatial domain technique in use. Despite the numerous benefits of RSMA from theoretical analysis, replacing SDMA with RSMA in practical systems will introduce several challenges. Although the \rev{SIC-free} receiver designs proposed in \cite{Sibo_TCOM} addressed the increase in receiver complexity, other challenges remain. These include:
\begin{itemize}
    \item Increased encoding/decoding complexity: with RSMA, more codewords are transmitted (especially with multi-layer schemes such as the hierarchical RSMA in \cite{Dai_HRS} and the general framework of multi-layer RSMA in \cite{Mao_EURASIP}), which leads to more complexity in encoding and decoding.
    \item Control signaling overhead: The increased number of streams requires an increased control signaling overhead to assist in decoding at receivers.
    \item Retransmission: As the common stream is intended for multiple users, it requires retransmission mechanisms that are different from SDMA, such as in \cite{Rafael}. The design of an efficient retransmission process for RSMA schemes poses a challenge.
\end{itemize}

Another direction that deserves exploration is to evaluate and design RSMA schemes using performance metrics that account for more practical limitations. The benefits of RSMA have been well studied under Gaussian input signals, for example, in \cite{Dai_HRS, Mao_EURASIP,Hamdi} and \cite{Li_JSAC}. The evaluation of achievable rates under finite block lengths is studied in \cite{Xu_FBL,Wang_FBL}. The achievable rates under finite-alphabet input constraints are studied in \cite{Sibo_TCOM,Sibo_SPAWC,Sibo_TWC}. 

The above works consider performance metrics originating from the mutual information, which is proved to be achievable with reliable communications by Shannon's channel coding theorem \cite{Shannon}. The proofs of the channel coding theorem assume optimal decoding, for example, the joint typicality decoding in \cite{Shannon} and the maximum-likelihood decoding in \cite{Gallager}. However, optimal decoding is often difficult in practice due to limited channel knowledge and/or complexity reasons. Therefore, it is interesting to discover achievable rates under suboptimal decoders, leading to the study of mismatched decoding \cite{Mismatched_1,Lapidoth_1,Albert_book}. A recent work \cite{Sibo_FAGCI} studies communication in finite-alphabet input channels under Gaussian noise and finite-alphabet interference, namely Finite-Alphabet Gaussian Channel under Interference (FAGCI), where the decoders fully or partially approximate the interference as Gaussian random variables to reduce computational complexity. This model reflects a common issue that interference is not properly detected in receivers for complexity reasons. A lower bound on the mismatched capacity is derived using the Generalized Mutual Information (GMI).

In this work, a novel downlink RSMA architecture is proposed, namely Codeword-Segmentation RSMA (CS-RSMA), to reduce the implementation challenges of RSMA. %We start by introducing the system model and principle of CS-RSMA, then apply the GMI derived in \cite{Sibo_FAGCI} for performance evaluation. Precoder optimization algorithms for Sum-Rate (SR) and Max-Min Fairness (MMF) optimization are also proposed for both conventional RSMA and CS-RSMA. We also propose a physical-layer implementation of CS-RSMA and evaluate it using link-level simulations.
The contribution of this work is summarized as follows:
\begin{itemize}
    \item We propose a novel downlink architecture for RSMA, namely CS-RSMA. Unlike conventional RSMA, which splits users' messages into common and private parts, then combines the common parts into a common message before encoding, CS-RSMA encodes users' messages directly, segments the resulting codewords into common and private parts, and combines the common parts into a common stream. Details on encoding and decoding procedures for both conventional RSMA and CS-RSMA are provided.
    
    \item We apply the GMI derived in \cite{Sibo_FAGCI} to analyze the achievable rate of CS-RSMA and conventional RSMA to capture receiver imperfection in practice. This leads to a different analysis from \cite{Sibo_TWC} where the achievable rate is measured by mutual information, which can only be achieved by optimal decoders, e.g., the maximum likelihood decoder. With the GMI, conventional RSMA and CS-RSMA can be evaluated under the assumption that the decoders treat undesired private streams as Gaussian random variables, despite the fact that they are from finite alphabets. Such suboptimal decoders reveal the performance of commonly used low-complexity receivers that treat interference as Gaussian noise instead of demodulating them, as demonstrated in \cite{Sibo_FAGCI}. To ensure a fair comparison, a decoding complexity constraint is defined.
    
    \item To reveal the theoretical performance of conventional RSMA and CS-RSMA, we develop precoder optimization methods to maximize Sum-Rate (SR) and Max-Min Fairness (MMF) among users, under the assumption of finite alphabets and suboptimal decoders. To exploit the flexibility of RSMA, we allow adaptive alphabet selection under a decoding complexity constraint. Numerical results from the proposed optimization algorithms are provided, which show that CS-RSMA outperforms conventional RSMA with and without SIC in terms of SR and performs similarly to the two in terms of MMF.
    
    \item While the achievable rate given by GMI measures theoretical limits and provides closed-form expressions for system design, they are based on idealized signal-space encoding and decoding, which remain a challenge in practice. Although the achievable rate of a proper channel model reflects the throughput in practical systems, a more accurate evaluation can be obtained through LLS, which involves practical designs, i.e., those incorporating binary encoding/decoding and modulation. For this purpose, as well as improving the practicality of CS-RSMA, we propose physical-layer (PHY) transceiver designs for CS-RSMA, and apply LLS to measure the Bit Error Rate (BER) performance of the proposed designs. Similarly to the achievable rate results, the LLS results show that CS-RSMA slightly outperforms conventional RSMA. From a practical perspective, the benefits of CS-RSMA over conventional RSMA are also discussed in terms of encoding/decoding complexity, control signalling overhead, and retransmission design.
\end{itemize}

\emph{Organization:} The rest of this paper is organized as follows. In Section II, we introduce the system model for both conventional RSMA and CS-RSMA. In Section III, results from \cite{Sibo_FAGCI} on multi-user communications with suboptimal decoders are briefly summarized. This then leads to the performance metric of conventional RSMA and CS-RSMA to be considered in this work, as well as a definition of decoding complexity. In Section IV, to evaluate the theoretical performance of conventional RSMA and CS-RSMA, SR and MMF optimization methods that capture the suboptimal decoders are proposed. Section V summarizes the numerical results on the theoretical performance. In Section VI, we propose the PHY implementation of CS-RSMA and evaluate it using LLS, and discuss the implementation benefits of CS-RSMA over conventional RSMA. Finally, Section VII concludes this paper.

\emph{Notations:} Scalars are denoted by normal letters.
Sequences and vectors are denoted by bold letters.
Random variables are denoted by capital letters, while the specific values are denoted by lowercase letters. For example, $x$ is an observation of $X$, and $\mathbf{x}$ is an observation of $\mathbf{X}$. Multisets are denoted by calligraphic letters. $P_{\cdot}(\cdot)$ and $P_{\cdot|\cdot}(\cdot|\cdot)$ denote probability and conditional probability respectively. $\mathbf{I}$ and $\mathbf{0}$ denote the identity matrix and the zero matrix, whose dimensions are given by their superscripts. $[\cdot]_{m,n}$ denotes the $(m,n)$-th entry of a matrix.  $(\cdot)^T$, $(\cdot)^H$, $\| \cdot \|$ and $\| \cdot \|_\text{F}$ denote respectively the transpose, conjugate transpose, Euclidean norm and Frobenius norm of the input entity. $|\cdot|$ denotes the absolute value if the argument is a scalar, or the cardinality if the argument is a set. $\times$ denotes the Cartesian product of two sets. $\mathbb{E}[\cdot]$ denotes the expectation. $\mathcal{CN}(\bm{\mu},\mathbf{\Sigma})$ denotes a circular symmetric complex Gaussian distribution for vectors whose mean and covariance matrix are $\bm{\mu}$ and $\mathbf{\Sigma}$. $\text{dim}(\cdot)$ denotes the dimension of the argument.

\section{Codeword Segmentation RSMA}

\begin{figure*}[t]
      \centering
      \includegraphics[width=6.5in]{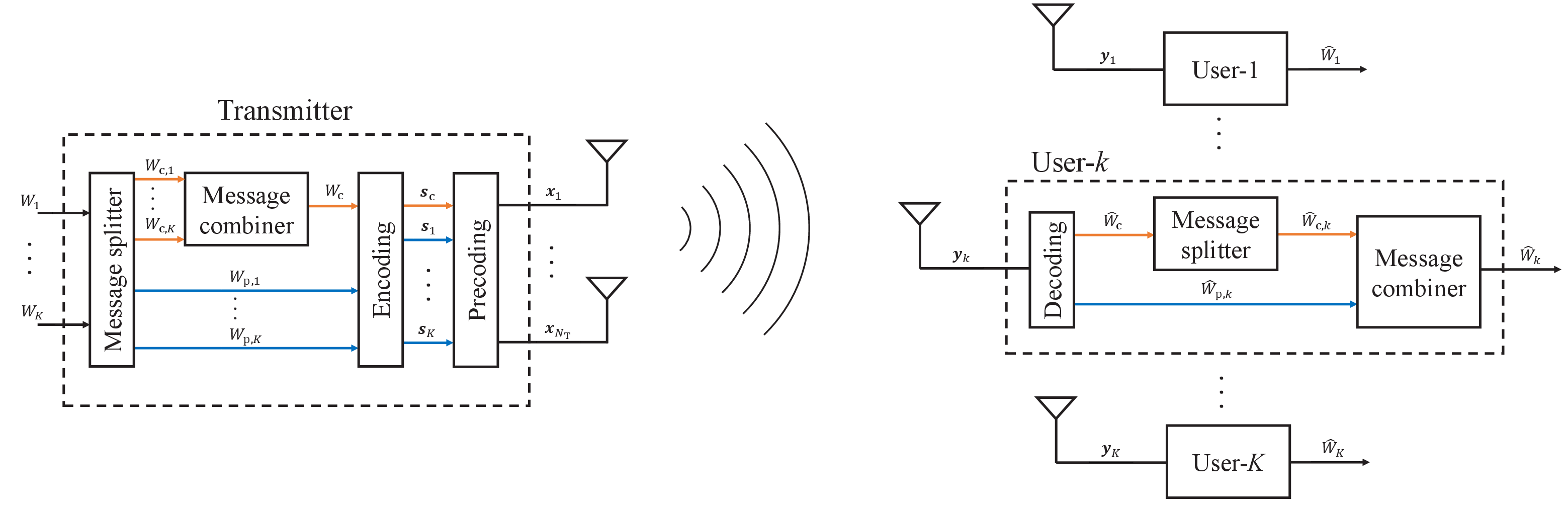}
      \caption{Conventional RSMA for multi-user MISO.}
      \label{Fig: conventional RSMA}

      \centering
      \includegraphics[width=6.5in]{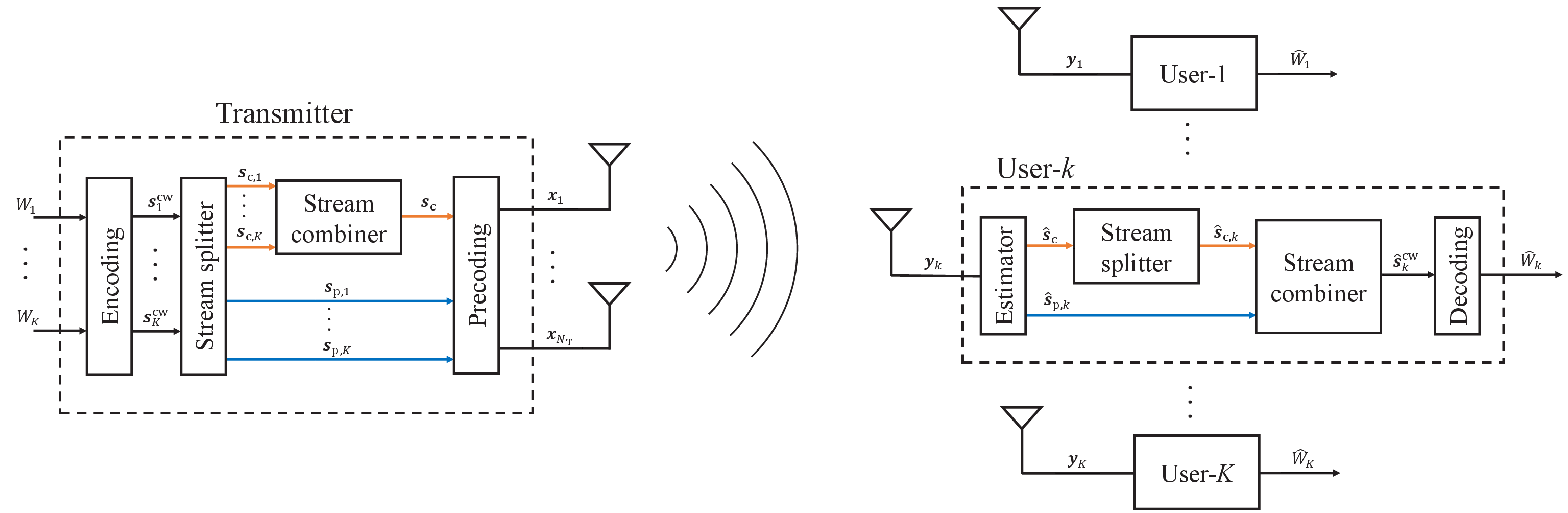}
      \caption{Codeword-Segmentation RSMA for multi-user MISO. }
      %\todo{Maybe later replace with tikz to maintain the same fonts.}
      \label{Fig: CS RSMA}
\end{figure*}

We consider a downlink multi-user Multiple-Input Single-Output (MISO) system where a transmitter equipped with $N_\text{T}$ transmit antennas serves $K$ users with unicast messages, $\{W_1,...,W_K\}$. We assume narrow-band channels and let $n$ denote the transmission block length. At time instant $t \in \{1,\;2,\;...,\;n\}$, the received signal at user-$k$, $k\in\mathcal{K}=\{1,\;...,\;K\}$, can be written as
\begin{equation}\label{equ: Rx signal}
    y_{t,k} = \mathbf{h}_{k}^H \mathbf{x}_t + z_{t,k},
\end{equation}
where $\mathbf{x}_t$ is the transmitted signal, $\mathbf{h}_{k}$ is the channel between the transmitter and user-$k$ and $z_{t,k}\sim\mathcal{CN}(0,\;\sigma_z^2)$ is the additive white Gaussian noise observed by user-$k$.
We assume that the transmitter applies single-layer RSMA \cite{Mao_EURASIP,Hamdi}, hence the transmit signal can be expressed as
\begin{equation}
    \mathbf{x}_t = \mathbf{P} \mathbf{s}_t,
\end{equation}
where $\mathbf{P} = [\mathbf{p}_{\text{c}},\; \;\mathbf{p}_{\text{p},1},\;...,\;\mathbf{p}_{\text{p},K}] \in\mathbb{C}^{N_\text{T} \times (K+1)}$ represents the precoding matrix and $\mathbf{s}_t = [s_{t,\text{c}} \;s_{t,\text{p},1},\;...\;,s_{t,\text{p},K}]^T \in \mathbb{C}^{(K+1) \times 1}$ represents the transmitted symbol vector. We assume $\mathsf{E}\{\mathbf{s}_t\mathbf{s}_t^H\} = \mathbf{I}$, and the transmission is subject to a power budget $P_\text{T}$, where $P_\text{T} =\mathbb{E}\{\mathbf{x}_t^H\mathbf{x}_t\} = \|\mathbf{P}\|_\text{F}^2$. A collection of symbol vectors at all the time instances can be written as $\mathbf{S} = [\mathbf{s}_1,\;\mathbf{s}_2,\;...,\;\mathbf{s}_n]\in\mathbb{C}^{(K+1)\times n}$. The rows of $\mathbf{S}$ are $K+1$ symbol blocks (streams), resulting in an alternative expression, $\mathbf{S} = [\mathbf{s}_\text{c},\;\mathbf{s}_{\text{p},1},\;...,\;\mathbf{s}_{\text{p},K}]^T$, where $\mathbf{s}_\text{c}$ and $\mathbf{s}_{\text{p},k}$, $\forall k\in\mathcal{K}$, are referred to as common and private streams in RSMA. We assume that the symbols in $\mathbf{s}_\text{c}$ and $\mathbf{s}_{\text{p},k}$, $\forall k \in \mathcal{K}$, are uniformly drawn from their corresponding alphabets, denoted by $\mathcal{X}_\text{c}$ and $\mathcal{X}_{\text{p},k}$.

\begin{figure}
    \centering  \includegraphics[width=1\linewidth]{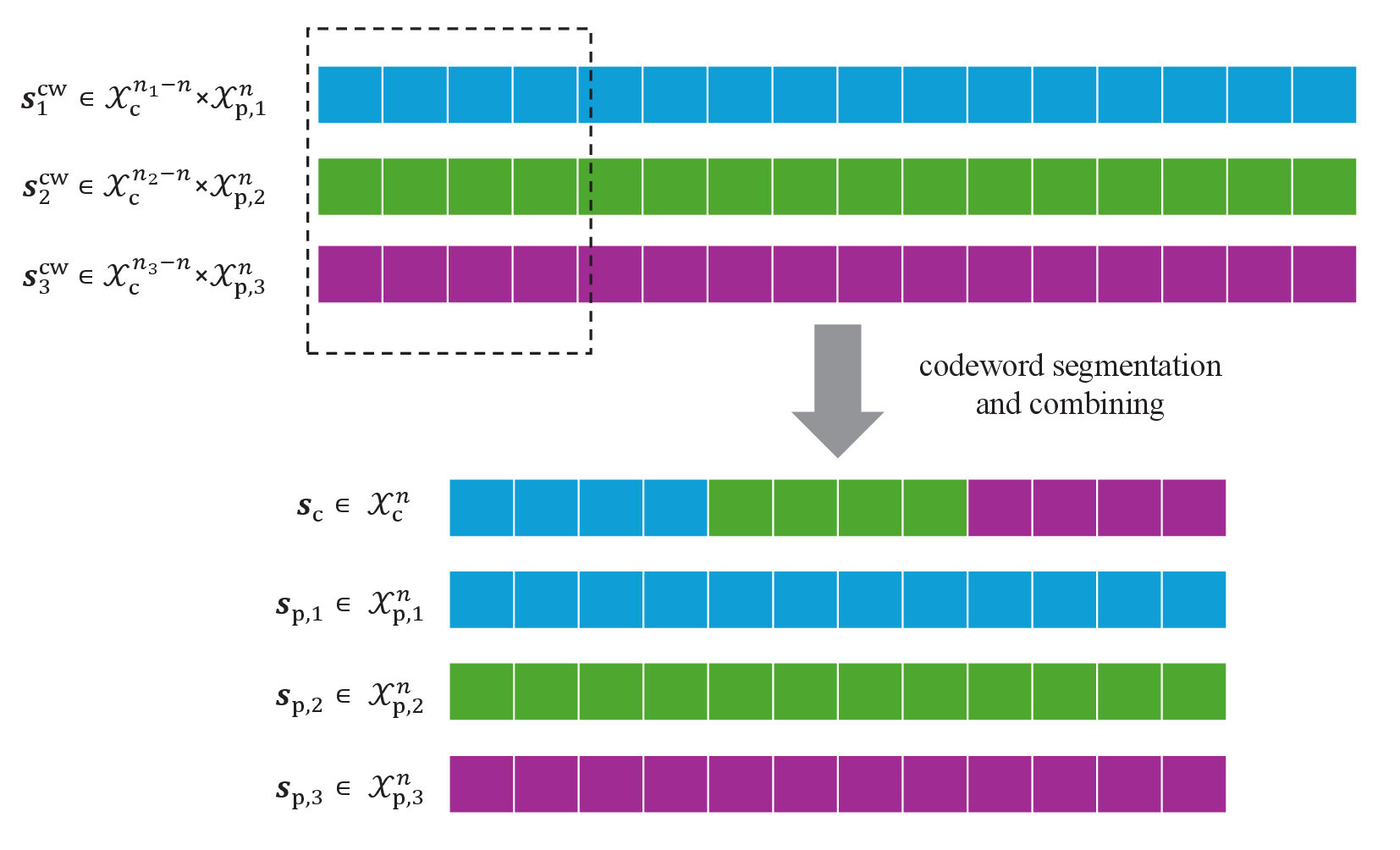}
    \caption{\rev{A visual illustration of codeword segmentation and combination.}}
    \label{fig: CS-RSMA block structure}
\end{figure}

The general procedures for encoding $\{W_1,...,W_K\}$ into $\{\mathbf{s}_{\text{c}},\;\mathbf{s}_{\text{p},1},\;...,\;\mathbf{s}_{\text{p},K}\}$ at the transmitter and decoding them in the receivers differ in conventional RSMA and the proposed CS-RSMA. Both are described below and are depicted in Fig. \ref{Fig: conventional RSMA} and \ref{Fig: CS RSMA}, without constraining the specific code and the decoding algorithm in use.
\begin{enumerate}
    \item \textit{Conventional RSMA:} $\{W_1,...,W_K\}$ are first split into common and private parts, $\{W_{\text{c},1},...,W_{\text{c},K}\}$ and $\{W_{\text{p},1},...,W_{\text{p},K}\}$, respectively. The common parts are combined into one common message, $W_{\text{c}}$, while the private parts remain separate as private messages. The $K+1$ messages, $\{W_{\text{c}},\;W_{\text{p},1},...,W_{\text{p},K}\}$, are then encoded into $K+1$ codewords, $\{\mathbf{s}_{\text{c}},\;\mathbf{s}_{\text{p},1},\;...,\;\mathbf{s}_{\text{p},K}\} \subset \mathbb{C}^{n\times 1}$. Upon reception, user-$k$ decodes the received signal into estimates of $W_\text{c}$ and $W_{\text{p},k}$, denoted by $\widehat{W}_\text{c}$ and $\widehat{W}_{\text{p},k}$. It then extracts from $\widehat{W}_\text{c}$ an estimate of its desired common part, denoted by $\widehat{W}_{\text{c},k}$. Finally, user-$k$ combines $\widehat{W}_{\text{c},k}$ and $\widehat{W}_{\text{p},k}$ to obtain an estimate of the desired unicast message, denoted by $\widehat{W}_k$.

    \item \textit{Codeword-Segmentation RSMA:} We introduce CS-RSMA as a novel alternative implementation to conventional RSMA. In CS-RSMA, $\{W_1,...,W_K\}$ are first encoded into $K$ codewords, $\{\mathbf{s}_1^\text{cw},\;...,\;\mathbf{s}_K^\text{cw}\}$, where $\mathbf{s}_k^\text{cw} \in \mathbb{C}^{n_k \times 1}$, $\forall k\in\mathcal{K}$, and $n_k$ is the codeword length for user-$k$ satisfying \rev{$n_k\geq n$} and the following constraint:
    \begin{equation}\label{equ: block_length_constraint}
        \sum_{k=1}^K n_k = (K+1)n.
    \end{equation}
    $\{\mathbf{s}_1^\text{cw},\;...,\;\mathbf{s}_K^\text{cw}\}$ are then segmented into private and common parts. The private parts have a uniform length $n$ and are denoted by $\{\mathbf{s}_{\text{p},1},\;...,\;\mathbf{s}_{\text{p},K}\}\subset\mathbb{C}^{n\times1}$. The common parts are denoted by $\{\mathbf{s}_{\text{c},1},\;...,\;\mathbf{s}_{\text{c},K}\}$, where $\mathbf{s}_{\text{c},k} \in \mathbb{C}^{(n_k-n) \times 1}$, $\forall k\in\mathcal{K}$. The common parts are then combined (e.g., concatenated or interleaved) into one symbol stream, denoted by $\mathbf{s}_\text{c}$, and the private parts remain as separate private streams. As a consequence of (\ref{equ: block_length_constraint}), $\mathbf{s}_\text{c}\in\mathbb{C}^{n\times1}$. \rev{An example of codeword segmentation and combination is illustrated in Fig. \ref{fig: CS-RSMA block structure}.} Upon reception, user-$k$ estimates $\mathbf{s}_\text{c}$ and $\mathbf{s}_{\text{p},k}$ from $y_k$ and output $\widehat{\mathbf{s}}_\text{c}$ and $\widehat{\mathbf{s}}_{\text{p},k}$. Then it extracts an estimate of its desired codeword segment from $\widehat{\mathbf{s}}_\text{c}$, denoted by $\widehat{\mathbf{s}}_{\text{c},k}$. $\widehat{\mathbf{s}}_{\text{c},k}$ and $\widehat{\mathbf{s}}_{\text{p},k}$ are combined into $\widehat{\mathbf{s}}_k^\text{cw}$, an estimate of $\mathbf{s}_k^\text{cw}$. Finally, based on $\widehat{\mathbf{s}}_k^\text{cw}$, user-$k$ decodes its desired message, $\widehat{W}_k$.
\end{enumerate}

The main difference between conventional RSMA and CS-RSMA lies in the order of encoding and splitting: at the transmitter, conventional RSMA splits the messages before encoding them, whereas CS-RSMA directly encodes the messages and segments the codewords; at the receivers, the conventional RSMA decodes the common and private streams before reconstructing the message, whereas CS-RSMA first reconstructs the desired codeword from the codeword segments, and then decodes the reconstructed codeword into the desired message. 

Regarding receiver implementation, SIC can be implemented with conventional RSMA. With SIC receivers, each user first decodes the common message, re-encode it into an estimate of the common stream, remove/cancel the estimated common stream from the received signal, and decode the desired private message. Meanwhile, \rev{SIC-free receivers, which do not rely on common stream decoding when decoding the private stream, provide a low-complexity option for conventional RSMA \cite{Sibo_TCOM}, since re-encoding and cancellation are not performed.} With CS-RSMA, SIC cannot be implemented. This is because the common stream is not a codeword, but a combination of segmented codewords from different users, and therefore cannot be decoded directly. \rev{The idea of SIC-free receivers is still valid for CS-RSMA.}\footnote{\rev{That said, due to the codeword segmentation, the receiver architecture of CS-RSMA is still different from SIC-free RSMA. For conventional RSMA, the SIC-free receiver decodes the common and private streams separately and combines the resulting messages. For CS-RSMA, the receiver processes the common and private symbol streams into the desired codeword and decodes it.}} Fortunately, \cite{Sibo_TWC} shows that \rev{SIC-free} receivers only incur a minor performance loss in RSMA compared to SIC receivers. Therefore, the absence of SIC in the CS-RSMA is not a critical limitation.

\section{Performance Evaluation using\\ Mismatched Decoding}
In \cite{Sibo_TCOM, Sibo_SPAWC, Sibo_TWC}, the performance of SIC-based and \rev{SIC-free} receivers is studied considering constellation-constrained mutual information. However, mutual information has only been shown to be achievable under the assumption of optimal decoding (e.g., the maximum likelihood decoding in \cite{Gallager}). Optimal decoders can be computationally prohibitive when the number of signals observed by the receivers increases. Recently, \cite{Sibo_FAGCI} investigated a channel model that takes finite-alphabet input and is disrupted by finite-alphabet interference and Gaussian noise, namely the FAGCI model. Under the FAGCI model, \cite{Sibo_FAGCI} considers suboptimal decoders that treat part of finite-alphabet interference as Gaussian random variables and derives the GMI as a performance metric under such suboptimal decoders. Strong connections between suboptimal decoders and demodulators are also depicted in \cite{Sibo_FAGCI}, making the GMI for the FAGCI model an accurate physical layer abstraction for formulating signal processing problems. Extension to a multiple-antenna setup is also discussed.

In this section, we first summarize the results on the GMI for the FAGCI model under MISO scenarios from \cite{Sibo_FAGCI}. Then, to address scenarios where complexity constraints are imposed on receivers, we apply the GMI to analyze and evaluate the performance of conventional RSMA and CS-RSMA.

\subsection{GMI for Finite Alphabet Input Gaussian Channel Under Interference}
A generic expression for the received signal at user-$k$, $\forall k \in \{1,\;...,\;K\}$, can be written as follows: 
\begin{equation}\label{equ: single-user MISO Rx}
    y_{t,k} = \mathbf{h}_k^H(\mathbf{p}_x x_{t,k} + \mathbf{P}_i\mathbf{i}_t + \mathbf{P}_j\mathbf{j}_t) + z_{t,k}.
\end{equation}
$x_{t,k}$, $\mathbf{i}_t$ and $\mathbf{j}_t$ are respectively the desired symbol for user-$k$, the interfering symbol vector to be treated optimally and the interfering symbol vector to be treated as Gaussian random variables by user-$k$, respectively drawn from $\mathcal{X}$, $\mathcal{I}$ and $\mathcal{J}$ with uniform distribution. $\mathbf{p}_x$, $\mathbf{P}_i$ and $\mathbf{P}_j$ are respectively the precoders for $x_t$, $\mathbf{i}_t$ and $\mathbf{j}_t$. $z_{t,k}$ is additive noise generated from the i.i.d. $\mathcal{CN}(0,\;\sigma_z^2)$. Without loss of generality, we assume that $\mathbb{E}[|x|^2] = 1$, $\mathbb{E}[\mathbf{i} \mathbf{i}^H] = \mathbf{I}^{\text{dim}(\mathbf{i})}$ and $\mathbb{E}[\mathbf{j} \mathbf{j}^H] = \mathbf{I}^{\text{dim}(\mathbf{j})}$ unless any of the alphabets is $\{0\}$, i.e., no signal is transmitted. The decoder at user-$k$ can be written as
\begin{equation}
    \hat{m} = \arg \max_{m=1,...,M} q^n(\mathbf{x}^{(m)}_k,\;\mathbf{y}_k),
\end{equation}
where $M$ is the size of the codebook, $\mathbf{x}^{(m)}_k$ is the $m$-th codeword, and $\mathbf{y}_k$ is the output sequence. \rev{We consider a memoryless decoding metric that can be decomposed as follows}:
\begin{equation}
    q^n(\mathbf{x}^{(m)}_k,\;\mathbf{y}_k) = \prod_{t=1}^n q(x^{(m)}_{t,k},\;y_{t,k}).
\end{equation}
\rev{This assumption is reasonable since the original channel in \eqref{equ: Rx signal} is memoryless.}

The optimal (matched) decoding metric is given by
\begin{equation}\label{equ: matched decoding metric}
    q(x,\;y) = \sum_{\Bar{\mathbf{i}}\in\mathcal{I}} \sum_{\Bar{\mathbf{j}}\in\mathcal{J}} \exp\left({-\frac{|y-\mathbf{h}_k^H\mathbf{P}_i\Bar{\mathbf{i}}-\mathbf{h}_k^H\mathbf{P}_j\Bar{\mathbf{j}}-\mathbf{h}_k^H\mathbf{p}_x x|^2}{\sigma_z^2}}\right).
\end{equation}
With fixed precoders, (\ref{equ: matched decoding metric}) achieves the matched capacity of (\ref{equ: single-user MISO Rx}), which is given by mutual information under finite-alphabet constraints.

It can be observed that the complexity of (\ref{equ: matched decoding metric}) depends on the cardinality of $\Bar{\mathbf{i}}$ and $\Bar{\mathbf{j}}$, which can be significant as the number of interference signals increases. In this work, we are interested in a suboptimal (mismatched) decoding metric given by
\begin{equation}\label{equ: mismatched decoding metric}
    q(x,\;y_k) = \sum_{\Bar{\mathbf{i}}\in\mathcal{I}} \exp\left({-\frac{|y_k-\mathbf{h}_k^H\mathbf{P}_i\Bar{\mathbf{i}}-\mathbf{h}_k^H\mathbf{p}_x x|^2}{\parallel\mathbf{h}_k^H\mathbf{P}_j\parallel^2 +\sigma_z^2}}\right).
\end{equation}
The philosophy of (\ref{equ: mismatched decoding metric}) is that the decoder uses full knowledge of $\mathbf{i}$ and treats $\mathbf{j}$ as Gaussian noise with the same variance. In fact, if $\mathbf{j} \sim \mathcal{CN}(\mathbf{0}^{\text{dim}(\mathbf{j})\times1},\;\mathbf{I}^{\text{dim}(\mathbf{j})})$, (\ref{equ: mismatched decoding metric}) is the optimal decoding metric for (\ref{equ: single-user MISO Rx}). It can be observed that the computational complexity of (\ref{equ: mismatched decoding metric}) depends only on $\mathbf{i}$ but not on $\mathbf{j}$, and is less than (\ref{equ: matched decoding metric}) by a factor of $|\mathcal{J}|$.

\begin{figure*}
\begin{flalign}\label{MISO GMI}
    &I_{\text{GMI},k}\left(\mathbf{h}_k^H\mathbf{p}_{x}\mathcal{X},\; \mathbf{h}_k^H\mathbf{P}_{i}\mathcal{I},\; \mathbf{h}_k^H\mathbf{P}_{j}\mathcal{J},\; \sigma_z^2\right) &&\nonumber\\
    =&\sup_{s\geq 0}\;\;\;\log|\mathcal{X}|-\frac{1}    {|\mathcal{X}||\mathcal{I}||\mathcal{J}|} \sum_{x\in\mathcal{X}}\sum_{\mathbf{i}\in\mathcal{I}}\sum_{\mathbf{j}\in\mathcal{J}}\mathbb{E}_Z\left[ \log \sum_{\Bar{x}\in\mathcal{X}} \left( \sum_{\Bar{\mathbf{i}}\in\mathcal{I}} \exp\left(-\frac{|\mathbf{h}_k^H\mathbf{p}_{x}(x-\Bar{x})+\mathbf{h}_k^H\mathbf{P}_{\mathbf{I}}(\mathbf{i}-\Bar{\mathbf{i}})+\mathbf{h}_k^H\mathbf{P}_{j}\mathbf{j}+z|^2}{\|\mathbf{h}_k^H\mathbf{P}_{j}\|^2 + \sigma_z^2}\right) \right)^s\right] &&\nonumber\\
    &+ \frac{1}{|\mathcal{I}||\mathcal{J}|} \sum_{\mathbf{i}\in\mathcal{I}}\sum_{\mathbf{j}\in\mathcal{J}} \mathbb{E}_Z\left[ \log \left(\sum_{\Bar{i}\in\mathcal{I}} \exp\left(-\frac{|\mathbf{h}_k^H\mathbf{P}_{i}(\mathbf{i}-\Bar{\mathbf{i}})+\mathbf{h}_k^H\mathbf{P}_{j}\mathbf{j}+z|^2}{\|\mathbf{h}_k^H\mathbf{P}_{j}\|^2 + \sigma_z^2}\right)^s \right) \right] &&
\end{flalign}
\begin{flalign}\label{MISO GMI approx}                      &I_{\text{GMI},k}^{\text{approx}}\left(\mathbf{h}_k^H\mathbf{p}_{x} \mathcal{X},\; \mathbf{h}_k^H\mathbf{P}_{i}\mathcal{I},\; \mathbf{h}_k^H\mathbf{P}_{j}\mathcal{J},\; \sigma_z^2\right) &&\nonumber\\ 
    =& \log|\mathcal{X}| - \frac{1}{|\mathcal{X}||\mathcal{I}||\mathcal{J}|} \sum_{x\in\mathcal{X}}\sum_{\mathbf{i}\in\mathcal{I}}\sum_{\mathbf{j}\in\mathcal{J}} \log \sum_{\Bar{x}\in\mathcal{X}} \sum_{\Bar{\mathbf{i}}\in\mathcal{I}} \exp\left(-\frac{|\mathbf{h}_k^H\mathbf{p}_{x}(x-\Bar{x})+\mathbf{h}_k^H\mathbf{P}_{i}(\mathbf{i}-\Bar{\mathbf{i}})+\mathbf{h}_k^H\mathbf{P}_{j}\mathbf{j}|^2}{\|\mathbf{h}_k^H\mathbf{P}_{j}\|^2+2\sigma_z^2}\right) &&\nonumber\\ 
    & + \frac{1}{|\mathcal{I}||\mathcal{J}|} \sum_{\mathbf{i}\in\mathcal{I}}\sum_{\mathbf{j}\in\mathcal{J}} \log \sum_{\Bar{\mathbf{i}}\in\mathcal{I}} \exp\left(-\frac{|\mathbf{h}_k^H\mathbf{P}_{i}(\mathbf{i}-\Bar{\mathbf{i}})+\mathbf{h}_k^H\mathbf{P}_{j}\mathbf{j}|^2}{\|\mathbf{h}_k^H\mathbf{P}_{j}\|^2+2\sigma_z^2}\right) &&
\end{flalign}
\begin{flalign}\label{equ: MISO GMI approx gradient}
    &\nabla_{\widetilde{\mathbf{P}}_k}
    I_{\text{GMI},k}^{\text{approx}}\left(\mathbf{h}_k^H\mathbf{p}_{x} \mathcal{X},\; \mathbf{h}_k^H\mathbf{P}_{i}\mathcal{I},\; \mathbf{h}_k^H\mathbf{P}_{j}\mathcal{J},\; \sigma_z^2\right) &&\nonumber\\
    =& -\frac{1}{|\mathcal{X}||\mathcal{I}||\mathcal{J}|} \sum_{x\in\mathcal{X}}\sum_{\mathbf{i}\in\mathcal{I}}\sum_{\mathbf{j}\in\mathcal{J}} \frac{1}{\sum_{\Bar{x}\in\mathcal{X}} \sum_{\Bar{\mathbf{i}}\in\mathcal{I}}g(\widetilde{\mathbf{P}}_k,\mathbf{v})}
    \sum_{\Bar{x}\in\mathcal{X}} \sum_{\Bar{\mathbf{i}}\in\mathcal{I}}g_1(\widetilde{\mathbf{P}}_k,\mathbf{v})
    + \frac{1}{|\mathcal{I}||\mathcal{J}|} \sum_{\mathbf{i}\in\mathcal{I}}\sum_{\mathbf{j}\in\mathcal{J}} \frac{1}{\sum_{\Bar{\mathbf{i}}\in\mathcal{I}} h(\widetilde{\mathbf{P}}_k,\mathbf{w})}
    \sum_{\Bar{\mathbf{i}}\in\mathcal{I}} h_1(\widetilde{\mathbf{P}}_k,\mathbf{w}) &&
\end{flalign}
\begin{flalign}
    \;\;\;\;\;\;\; &g(\widetilde{\mathbf{P}}_k,\;\mathbf{v}) = \exp\left(-\frac{|\mathbf{h}_k^H\widetilde{\mathbf{P}}_k\mathbf{v}|^2}{\|\mathbf{h}_k^H\widetilde{\mathbf{P}}_k\;\mathbf{I}_{j}\|^2 + 2\sigma_z^2}\right) &&\label{g(Pv)}\\ 
    &h(\widetilde{\mathbf{P}}_k,\;\mathbf{w}) = \exp\left(-\frac{|\mathbf{h}_k^H\widetilde{\mathbf{P}}_k\mathbf{w}|^2}{\|\mathbf{h}_k^H\widetilde{\mathbf{P}}_k\;\mathbf{I}_{j}\|^2 + 2\sigma_z^2}\right) &&\label{h(Pv)}\\
    &g_1(\widetilde{\mathbf{P}}_k,\;\mathbf{v}) = g(\widetilde{\mathbf{P}}_k,\;\mathbf{v})\left(\frac{-\left(\|\mathbf{h}_k^H\widetilde{\mathbf{P}}_k\;\mathbf{I}_{j}\|^2 + 2\sigma_z^2\right)\mathbf{h}_k\mathbf{h}_k^H\widetilde{\mathbf{P}}_k\mathbf{v}\mathbf{v}^H + (\mathbf{v}^H\widetilde{\mathbf{P}}_k^H\mathbf{h}_k\mathbf{h}_k^H\widetilde{\mathbf{P}}_k\mathbf{v}) \mathbf{h}_k\mathbf{h}_k^H\widetilde{\mathbf{P}}_k\;\mathbf{I}_{j}\;\mathbf{I}_{j}^H}{\left(\|\mathbf{h}_k^H\widetilde{\mathbf{P}}_k\;\mathbf{I}_{j}\|^2 + 2\sigma_z^2\right)^2}\right) &&\label{g1(Pv)}\\
    &h_1(\widetilde{\mathbf{P}}_k,\;\mathbf{w}) = h(\widetilde{\mathbf{P}}_k,\;\mathbf{w})\left(\frac{-\left(\|\mathbf{h}_k^H\widetilde{\mathbf{P}}_k\;\mathbf{I}_{j}\|^2 + 2\sigma_z^2\right)\mathbf{h}_k\mathbf{h}_k^H\widetilde{\mathbf{P}}_k\mathbf{w}\mathbf{w}^H + (\mathbf{w}^H\widetilde{\mathbf{P}}_k^H\mathbf{h}_k\mathbf{h}_k^H\widetilde{\mathbf{P}}_k\mathbf{w}) \mathbf{h}_k\mathbf{h}_k^H\widetilde{\mathbf{P}}_k\;\mathbf{I}_{j}\;\mathbf{I}_{j}^H}{\left(\|\mathbf{h}_k^H\widetilde{\mathbf{P}}_k\;\mathbf{I}_{j}\|^2 + 2\sigma_z^2\right)^2}\right) &&\label{h1(Pv)}
\end{flalign}
\hrule
\end{figure*}

For mismatched decoding problems, the GMI gives the highest achievable rate under the i.i.d. random coding ensemble \cite{Albert_book}. For the decoding metric given by (\ref{equ: mismatched decoding metric}), the GMI is given by (\ref{MISO GMI}) and is approximated by (\ref{MISO GMI approx}). \footnote{\rev{In \eqref{MISO GMI}, the optimization over $s$ is a convex problem, and therefore can be solved using gradient descent or bisection.}} To enable optimization of GMI, the gradient of GMI w.r.t. precoders is given by (\ref{equ: MISO GMI approx gradient}), where $g(\widetilde{\mathbf{P}}_k,\mathbf{v})$, $h(\widetilde{\mathbf{P}}_k,\mathbf{v})$, $g_1(\widetilde{\mathbf{P}}_k,\mathbf{v})$ and $h_1(\widetilde{\mathbf{P}}_k,\mathbf{v})$ are given by (\ref{g(Pv)}-\ref{h1(Pv)}), with 
\begin{equation}
\widetilde{\mathbf{P}}_k = [\mathbf{p}_x\;\; \mathbf{P}_i\;\; \mathbf{P}_j],
\end{equation}
\begin{equation}
\mathbf{v} = [x-\Bar{x};\;\;\;\; \mathbf{i}-\Bar{\mathbf{i}};\;\;\;\; \mathbf{j}],
\end{equation}
\begin{equation}
\mathbf{w} = [\mathbf{0}^{\text{dim}(\mathcal{X})\times 1};\;\;\;\; \mathbf{i}-\Bar{\mathbf{i}};\;\;\;\; \mathbf{j}],
\end{equation}
\begin{equation}
\mathbf{I}_j= [\mathbf{0}^{\text{dim}(\mathcal{X})\times \text{dim}(\mathcal{J})};\;\;\;\; \mathbf{0}^{\text{dim}(\mathcal{I})\times \text{dim}(\mathcal{J})};\;\;\;\; \mathbf{I}^{\text{dim}(\mathcal{J})}].
\end{equation}
Note that $\widetilde{\mathbf{P}}_k$ is a column-wise permutation of the precoding matrix at the transmitter based on user-$k$'s decoding strategy.

\subsection{Evaluation of RSMA and CS-RSMA}\label{subsec: decoding strategy and GMI}

We assume that each user applies a suboptimal decoding strategy by treating the desired signals optimally, but the undesired signals as Gaussian random variables with the same variance.\footnote{This is not the only possible decoding strategy. For clarity, we leave the analysis of other decoding strategies for future work.} For example, when user-$k$ decodes $s_\text{c}$, it treats $\mathbf{h}_k^H \mathbf{p}_k s_{\text{p},k}$ optimally, but treats $\{\mathbf{h}_k^H \mathbf{p}_{k'} s_{k'} \}$ as $\mathcal{CN}(0,|\mathbf{h}_k^H\mathbf{p}_{k'}|^2)$, $\forall k' \in \mathcal{K}/k$. This decoding strategy is based on the assumption that desired signals are typically stronger than unwanted signals. An implication of \cite{Sibo_FAGCI} is that treating strong interference as Gaussian random variables can lead to significant losses, whereas the impact of treating weak interference as Gaussian random variables can be negligible. Hence, we consider decoders that treat the desired signals optimally to enable high performance under limited allowance for decoding complexity.

\rev{Next, we present instantaneous information rate expressions associated with the common and private streams in relation to user-$k$. Shortly, we will see how these expressions are used to formulate provably achievable rates. These are given as follows:}
\begin{itemize}
    \item For the common stream:
        \begin{equation}\label{equ: common rate}
            I_{\text{c},k} = I_\text{GMI}\left(\mathbf{h}_k^H\mathbf{p}_\text{c}\mathcal{X}_\text{c},\; \mathbf{h}_k^H\mathbf{p}_k\mathcal{X}_{\text{p},k},\; \sum_{k'\neq k}\mathbf{h}_k^H\mathbf{p}_{k'}\mathcal{X}_{\text{p},k'},\; \sigma_z^2\right).
        \end{equation}
    \item For the private stream assuming SIC is applied to remove the common stream signal:
        \begin{equation}\label{equ: private rate SIC}
            I_{\text{p},k}^{\text{SIC}} = I_\text{GMI}\left(\mathbf{h}_k^H\mathbf{p}_k\mathcal{X}_{\text{p},k},\; \{0\},\; \sum_{k'\neq k}\mathbf{h}_k^H\mathbf{p}_{k'}\mathcal{X}_{\text{p},k'},\; \sigma_z^2\right).
        \end{equation}
    \item For the private stream assuming \rev{SIC-free} receivers, i.e., the decoding of the private stream is under interference from the common stream:
        \begin{equation}\label{equ: private rate non-SIC}
            I_{\text{p},k}^{\text{\rev{SIC-free}}} = I_\text{GMI}\left(\mathbf{h}_k^H\mathbf{p}_k\mathcal{X}_{\text{p},k},\; \mathbf{h}_k^H\mathbf{p}_\text{c}\mathcal{X}_\text{c},\; \sum_{k'\neq k}\mathbf{h}_k^H\mathbf{p}_{k'}\mathcal{X}_{\text{p},k'},\; \sigma_z^2\right).
        \end{equation}
\end{itemize}

$\forall k\in\mathcal{K}$, let $c_k$ denote the portion of $W_\text{c}$ occupied by $W_{\text{c},k}$ in conventional RSMA, or the portion of $\mathbf{s}_\text{c}$ occupied by $\mathbf{s}_{\text{c},k}$ in CS-RSMA. \rev{Hence, $c_k = (n_k-n)/n$ for CS-RSMA.} Therefore, $c_k\geq0$ and $\sum_{k\in\mathcal{K}} c_k= 1$. 

\rev{First, and for the sake of comparison, we present the rates achievable by user-$k$  under conventional RSMA. There are given by:}
\begin{itemize}
    \item Conventional RSMA with SIC:
        \begin{equation}\label{equ: rate RSMA SIC}
            I_{\text{GMI},k} = c_k\; \min_{k' \in \mathcal{K}}\; I_{\text{c},k'} + I_{\text{p},k}^{\text{SIC}}.
        \end{equation}
    \item Conventional RSMA with \rev{SIC-free} receivers:
        \begin{equation}\label{equ: rate RSMA SIC-free}
            I_{\text{GMI},k} = c_k\; \min_{k' \in \mathcal{K}}\; I_{\text{c},k'} + I_{\text{p},k}^{\text{\rev{SIC-free}}}.
        \end{equation}
\end{itemize}
\begin{proposition}[\rev{Achievable rate of CS-RSMA}]
    \rev{With a decoding metric that is separable in $s_{t,\text{c}}$ and $s_{t,\text{p},k}$, the achievable rate by user-$k$ under CS-RSMA is given by
    \begin{equation}\label{equ: rate CS-RSMA}
            I_{\text{GMI},k} = c_k I_{\text{c},k} + I_{\text{p},k}^{\text{SIC-free}}.
    \end{equation}}
\end{proposition}
\begin{IEEEproof}
    \rev{See Appendix \ref{appdx: achievability proof}.}
\end{IEEEproof} 

\begin{remark}
    The ``$\min$" operator in (\ref{equ: rate RSMA SIC}) and (\ref{equ: rate RSMA SIC-free}) is due to the fact that conventional RSMA requires all users to be able to decode the common stream. For CS-RSMA, users only need to decode the codeword after combining the segments from the common and private streams, and do not need to decode the common stream. \rev{The achievable rate at user-$k$ is therefore given by a weighted average of the instantaneous information rate of the common stream and the private stream, with the weights determined by the ratio of symbols assigned to the two streams.} Hence, the ``$\min$" operator disappears in (\ref{equ: rate CS-RSMA}).
\end{remark} 

\begin{remark}
    The common stream in CS-RSMA is a combination of codeword segments from potentially all the users, and is not meant to be decodable to any of the users. Therefore, applying SIC to decode and remove the common stream is not feasible for CS-RSMA. Since the private stream is disrupted by interference from the common stream in decoding, in (\ref{equ: rate CS-RSMA}), \rev{the information rate contributed by} the private stream is the same as in conventional \rev{SIC-free} RSMA, assuming that the decoding metrics for the common and desired private streams are separable \rev{(see Appendix \ref{appdx: achievability proof})}.
\end{remark}
\begin{remark}
    \rev{By comparing \eqref{equ: rate RSMA SIC-free} and \eqref{equ: rate CS-RSMA}, it can be concluded that the achievable rate of CS-RSMA is lower bounded by that of conventional SIC-free RSMA. This conclusion is independent of the decoding metric and can be generalized to GMI metrics based on any other decoding metric, including the mutual information, which is a special case of the GMI. This is because, in Appendix B, the proof does not assume any specific decoding metric, except its separability in common and private streams.}
\end{remark}

\subsection{Constrained Decoding Complexity}
Comparing communication schemes in the finite alphabet regime is more involved than in the Gaussian signaling regime. On the one hand, the achievable rate ultimately depends on the cardinality of the alphabets under sufficient SNR. On the other hand, the decoding complexity is proportional to the number of observed signals and the cardinality of their alphabets, which can be observed from (\ref{equ: matched decoding metric}) and (\ref{equ: mismatched decoding metric}). Therefore, we introduce the concept of decoding complexity constraint to ensure fairness of comparison. 

\begin{definition}[Decoding Complexity]
    A decoder of complexity-$\delta$ can compute ``$\exp(|\cdot|^2)$" for at most $n$ times for each received symbol.
\end{definition}

For example, for the channel model given by (\ref{equ: single-user MISO Rx}), the matched decoding, as in (\ref{equ: matched decoding metric}), is only valid if $\delta \geq |\mathcal{X}||\mathcal{I}||\mathcal{J}|$. The mismatched decoder given by (\ref{equ: mismatched decoding metric}) is valid if $\delta \geq |\mathcal{X}||\mathcal{I}|$, therefore, it requires a lower decoding complexity than the matched decoding.

\begin{proposition}[Decoding Complexity of RSMA]
    For conventional RSMA (with or without SIC) and CS-RSMA, if user-$k$ treats all the undesired private streams as Gaussian noise, the decoding is feasible with at least complexity-$|\mathcal{X}_\text{c}||\mathcal{X}_{\text{p},k}|$ at user-$k$, $\forall k \in \mathcal{K}$.
\end{proposition}

\begin{IEEEproof}
For all the RSMA schemes considered, we show that, if user-$k$ treats all the undesired private streams as Gaussian random variables, it needs to compute 
\begin{equation}\label{equ: RSMA decoding element}
    d(s_\text{c},\;s_{\text{p},k},\;y_{t,k}) = \exp\left(-\frac{|y_{t,k}-\mathbf{h}_k^H\mathbf{p}_{\text{c}}s_\text{c}-\mathbf{h}_k^H\mathbf{p}_k s_{\text{p},k}|^2}{\sum_{k'\neq k}|\mathbf{h}_k^H\mathbf{p}_{k'}|^2+\sigma_z^2}\right),
\end{equation}
 $\forall s_\text{c} \in \mathcal{X}_\text{c}$, $\forall s_{\text{p},k} \in \mathcal{X}_{\text{p},k}$ and for each time instance $t$. The rest of the decoding is achieved by only summing and multiplying $d(s_\text{c},\; s_{\text{p},k},\; y_{t,k})$ with different arguments. The exact decoders in terms of $d(s_\text{c},\;s_{\text{p},k},\;y_{t,k})$ are listed as follows: 
\begin{itemize}
    \item For conventional RSMA, Let $\{\mathbf{s}_\text{c}^{(1)},\;...,\;\mathbf{s}_\text{c}^{(M_\text{c})}\}$ denote the codebook for common stream and $\{\mathbf{s}_{\text{p},k}^{(1)},\;...,\;\mathbf{s}_{\text{p},k}^{(M_{\text{p},k})}\}$ denote the codebook for the private stream intended for user-$k$, $\forall k \in \mathcal{K}$, with $M_\text{c}$ and $M_{\text{p},k}$ being the corresponding codebook sizes. Let $s_{\text{c},t}^{(q)}$ and $s_{\text{p},t,k}^{(q)}$ respectively denote the symbol of $\mathbf{s}_\text{c}^{(q)}$ and $\mathbf{s}_{\text{p},k}^{(q)}$ at time instance $t$. The decoder for the common stream can be represented as
        \begin{equation}\label{equ: RSMA SIC decoding}
        \begin{split}
            \widehat{W}_\text{c} 
            =& \arg \max_{q\in{1,\;...,\;M_\text{c}}} q^n(\mathbf{s}_\text{c}^{(q)},\;\mathbf{y}_k)\\
            =& \arg \max_{q\in{1,\;...,\;M_\text{c}}} \prod_{t=1}^n \sum_{s_{\text{p},k}\in\mathcal{X}_{\text{p},k}} d(s_{\text{c},t}^{(q)},\;s_{\text{p},k},\;y_{t,k}).
        \end{split}
        \end{equation}
    If SIC is applied, the decoder for the private stream is based on $\widehat{W}_\text{c}$ and can be represented as
        \begin{equation}\label{equ: RSMA non-SIC decoding}
        \begin{split}
            \widehat{W}_k 
            =& \arg \max_{q\in{1,\;...,\;M_{\text{p},k}}} q^n(\mathbf{s}_{\text{p},k,t}^{(q)},\;\mathbf{y}_k)\\
            =& \arg \max_{q\in{1,\;...,\;M_{\text{p},k}}} \prod_{t=1}^n \; d(s_{\text{c},t}^{(\widehat{W}_\text{c})},\;s_{\text{p},t,k}^{(q)},\;y_{t,k}).
        \end{split}
        \end{equation}  
    If SIC is not applied,
        \begin{equation}\label{equ: CS-RSMA decoding}
        \begin{split}
            \widehat{W}_k 
            =& \arg \max_{q\in{1,\;...,\;M_k}} q^n(\mathbf{s}_{\text{p},k}^{(q)},\;\mathbf{y}_k)\\
            =& \arg \max_{q\in{1,\;...,\;M_k}} \prod_{t=1}^n \sum_{s_\text{c}\in\mathcal{X}_\text{c}} d(s_\text{c},\;s_{\text{p},t,k}^{(q)},\;y_{t,k}).
        \end{split}
        \end{equation}
    \item For CS-RSMA, let $\{\mathbf{s}_k^{(1)},\;...,\;\mathbf{s}_k^{(M_k)}\}$ represent the codebook for user-$k$ with $M_k$ being the codebook size, $\forall k \in \mathcal{K}$, and $s^{(q)}_{t,k}$ represent the symbol of $q$-th codeword at time instance $t$. The decoder can be represented as
        \begin{equation}
        \begin{split}
            \widehat{W}_k
            = \arg \max_{q\in{1,\;...,\;M_k}} &q^n(\mathbf{s}_k^{(q)},\;\mathbf{y}_k)\\
            = \arg \max_{q\in{1,\;...,\;M_k}} &\prod_{t'\in \mathcal{T}_k} \sum_{s_{k}\in\mathcal{X}_{\text{p},k}} d(s_{\tau_\text{c}(t'),k}^{(q)},\;s_{k},\;y_{t,k})\\
            &\prod_{t=1}^{n} \: \sum_{s_\text{c}\in\mathcal{X}_\text{c}} d(s_\text{c},\;s_{\tau_{\text{p},k}(t),k}^{(q)},\;y_{t,k}),
        \end{split}
        \end{equation}
    where $\mathcal{T}_k$ represents the time indices where the symbols of $\mathbf{s}_\text{c}$ are generated from $\mathbf{s}_{\text{c},k}$, and $\tau_\text{c}(\cdot)$ and $\tau_{\text{p},k}(\cdot)$ respectively maps the time index into the corresponding symbol index of user-$k$'s codeword.
\end{itemize}
Hence, computing $d(s_\text{c},\;s_{\text{p},k},\;y_{t,k})$, $\forall s_\text{c} \in \mathcal{X}_\text{c}$ and $\forall s_{\text{p},k} \in \mathcal{X}_{\text{p},k}$, covers all ``$\exp(|\cdot|^2)$" operations required for (\ref{equ: RSMA SIC decoding})-(\ref{equ: CS-RSMA decoding}). This proves that complexity-$\delta$ is sufficient to decode conventional RSMA and CS-RSMA. It is also necessary to compute $d(s_\text{c},\;s_{\text{p},k},\;y_{t,k})$, $\forall s_\text{c} \in \mathcal{X}_\text{c}$ and $\forall s_{\text{p},k} \in \mathcal{X}_{\text{p},k}$, as long as, at each time instance and for each stream, the probability of any symbol in the alphabet occurring is non-zero.
\end{IEEEproof}

\section{Precoder Optimization}\label{sec: Opt algo}
We propose precoder optimization algorithms for SR and MMF optimization under finite alphabets and suboptimal decoders. The alphabets are assumed to be fixed in this section, but will be allowed to adjust adaptively in Section \ref{Sec: Opt results}. To reduce computational complexity, the approximation given by (\ref{MISO GMI approx}) is utilized in the optimization, but the performance evaluation in Section \ref{Sec: Opt results} will be based on the exact GMI given by (\ref{MISO GMI}).

\subsection{Sum-Rate Optimization}

We consider a sum-rate maximization problem as follows:
\begin{equation}
\begin{split}\label{equ: SR initial formulation}
    \mathcal{P}_1: \;\; \underset{\mathbf{P},\mathbf{c}}{\max} \: \: & \sum_{k\in\mathcal{K}} I_{\text{GMI},k}\\ 
    \text{s.t.} \: \: \: \;
    & \|\mathbf{P}\|_\text{F}^2 \leq P_\text{T}\\
    & \mathbf{c}^T \mathbf{1} = 1\\
    & \mathbf{c} \succcurlyeq \mathbf{0},
\end{split}
\end{equation}
where $\mathbf{c} = [c_1,\;c_2,\;...,\;c_K]^T$.

\begin{proposition}[Sum-rate maximization reformulation]\label{SR max reformulation}
    Under different RSMA schemes, $\mathcal{P}_1$ can be simplified into the following formulations:
\begin{itemize}
    \item For conventional RSMA with or without SIC,
        \begin{equation}\label{equ: SR reformulation RSMA}
        \begin{split}
            \mathcal{P}_2: \;\; \underset{\mathbf{P}}{\max} \: \: 
            & \min_{k\in\mathcal{K}}\; I_{\text{c},k} + \sum_{k\in\mathcal{K}} I_{\text{p},k}^{\text{SIC/\rev{SIC-free}}}\\ 
            \text{s.t.} \: \: \: \;
            & \|\mathbf{P}\|_\text{F}^2 \leq P_\text{T}.\\
        \end{split}
        \end{equation}
    \item For CS-RSMA,
        \begin{equation}\label{equ: SR reformulation CS-RSMA}
        \begin{split}
            \mathcal{P}_3: \;\; \underset{\mathbf{P}}{\max} \: \: 
            & \max_{k\in\mathcal{K}}\; I_{\text{c},k} + \sum_{k\in\mathcal{K}} I_{\text{p},k}^{\text{\rev{SIC-free}}}\\ 
            \text{s.t.} \: \: \: \;
            & \|\mathbf{P}\|_\text{F}^2 \leq P_\text{T}.\\
        \end{split}
        \end{equation}
\end{itemize}
\end{proposition}

\begin{IEEEproof}
    (\ref{equ: SR reformulation RSMA}) is readily seen by substituting (\ref{equ: rate RSMA SIC}) and (\ref{equ: rate RSMA SIC-free}) into (\ref{equ: SR initial formulation}). For (\ref{equ: SR reformulation CS-RSMA}), substituting (\ref{equ: rate CS-RSMA}) into (\ref{equ: SR initial formulation}) leads to the following objective function:
    \begin{equation}
        \sum_{k\in\mathcal{K}} c_k I_{\text{c},k} + \sum_{k\in\mathcal{K}} I_{\text{p},k}^{\text{\rev{SIC-free}}}.
    \end{equation}
    It is trivial to see that the optimal $\mathbf{c}$ is always given by having $c_k = 1$ for $k = \arg\max_k I_{\text{c},k}$, and $c_{k'} = 0$ for $k' \neq k$. Using this optimal $\mathbf{c}$ results in (\ref{equ: SR reformulation CS-RSMA}).
\end{IEEEproof}

\begin{remark}\label{Remark: SR min to max}
    An interesting difference between $\mathcal{P}_2$ and $\mathcal{P}_3$ is that the ``$\min$" operator in $\mathcal{P}_2$ is replaced by ``$\max$" in $\mathcal{P}_3$. This indicates that CS-RSMA is expected to perform at least as well as conventional RSMA without SIC in SR, since its problem formulation only differs from the latter by changing the ``$\min$" to a ``$\max$". The comparison between CS-RSMA and conventional RSMA with SIC is difficult to reveal analytically because, although CS-RSMA is disadvantaged in the omission of SIC, its common stream is not restricted to be decodable by all the users. 
\end{remark}

\begin{remark}
    To achieve the maximum SR of CS-RSMA, if the common stream is activated, the number of codewords to split is strictly one. In contrast, in conventional RSMA, the number of messages to split can be any number between 1 and $K$, as discussed in \cite{Sibo_TWC}.
\end{remark}

We further reformulate the problems in Proposition \ref{SR max reformulation} using the logarithmic barrier \cite{boyd2004convex} to approximate $\mathcal{P}_2$ and $\mathcal{P}_3$ as the following unconstrained optimization problem:
\begin{equation}\label{SR_unconstrained_form}
\begin{split}
    \mathcal{P}_4: \;\; \underset{\mathbf{P},\mathbf{c}}{\max} \;\; 
    & \tau\:I_{\text{sum}} + \log \left(P_\text{T} - \parallel\mathbf{P}\parallel_\text{F}^2\right),\\ 
\end{split}
\end{equation}
where $I_\text{sum}$ represents the objective function of $\mathcal{P}_2$ and $\mathcal{P}_3$ depending on the RSMA scheme in use, and $\tau$ is a parameter that sets the accuracy of the approximation.

For RSMA with SIC, one subgradient of $I_\text{sum}$ w.r.t. $\mathbf{P}$ is given by
\begin{equation}\label{equ: gradient SIC}
    g(I_\text{sum}) = 
    \nabla_\mathbf{P} I_{\text{c},k'} + \sum_{k\in\mathcal{K}} \nabla_\mathbf{P} I_{\text{p},k}^{\text{SIC}},
\end{equation}
where $k'= \arg \min_k I_{\text{c},k}$. For RSMA without SIC and CS-RSMA, 
\begin{equation}\label{equ: gradient CS-RSMA/non-SIC}
    g(I_\text{sum}) = 
    \nabla_\mathbf{P} I_{\text{c},k'} + \sum_{k\in\mathcal{K}} \nabla_\mathbf{P} I_{\text{p},k}^{\text{\rev{SIC-free}}},
\end{equation}
where, for RSMA without SIC, $k'= \arg \min_k I_{\text{c},k}$ and, for CS-RSMA, $k'= \arg \max_k I_{\text{c},k}$. The gradients in (\ref{equ: gradient SIC}) and (\ref{equ: gradient CS-RSMA/non-SIC}) can be computed by referring to (\ref{equ: common rate})-(\ref{equ: private rate non-SIC}) and (\ref{equ: MISO GMI approx gradient}). 

It can be noticed that the problems described by $\mathcal{P}_1-\mathcal{P}_4$ are non-convex due to the non-convexity of (\ref{MISO GMI}) and (\ref{MISO GMI approx}) w.r.t. the precoders, and non-smooth due to the ``$\min$/$\max$" operators over $k$. Hence, it is difficult to obtain the global optimal solutions of them. We use the subgradient descent method to optimize $\mathcal{P}_4$. The complete algorithm is summarized in Alg. \ref{alg: GD for SR opt}, with $\Omega_\text{SR}(\mathbf{P})$ representing the objective function in $\mathcal{P}_4$. Alg. \ref{alg: GD for SR opt} is guaranteed to converge because each update on $\mathbf{P}$ leads to a non-decreasing update on the objective function, and the objective function is clearly upper-bounded.

\begin{algorithm}[!t]
    \caption{Subgradient Ascent with Barrier Method for Sum-Rate Maximization}
    \label{alg: GD for SR opt}
    \begin{algorithmic}[1]
        \REQUIRE $P_\text{T}$, $\mathcal{X}_\text{c}$, $\mathcal{X}_{\text{p},k}$, $\mathbf{h}_k$, $k\in\mathcal{K}$, $\beta$, $\epsilon$, $v_\text{max}$, $\tau_\text{max}$.
        \ENSURE $\mathbf{P}^\star$
        \STATE Initialize $\mathbf{P}^0$ and $\tau$.
        \REPEAT 
            \STATE $v:=0$.
            \REPEAT
                \STATE $\Delta\mathbf{P} := \tau\:g( I_\text{sum}) + \frac{\mathbf{P}^v}{\parallel \mathbf{P}^v \parallel_\text{F}^2 - P_\text{T}}$.
                \STATE Choose step size $\alpha$ via backtracking line search.
                \STATE Update. $\mathbf{P}^{v+1} := \mathbf{P}^v + \alpha \Delta\mathbf{P}$.
                \STATE Update. $\Omega_\text{SR}^{v+1} := \Omega_\text{SR}(\mathbf{P}^{v+1})$.
                \STATE $v:=v+1$.
            \UNTIL $\mid \Omega_\text{SR}^{v} - \Omega_\text{SR}^{v-1} \mid\; < \epsilon$ or $v \geq v_\text{max}$.
            \STATE Update $\tau := \beta \tau$.
        \UNTIL $\tau \geq \tau_\text{max}$.
        \RETURN $\mathbf{P}^\star := \mathbf{P}^{v}$.
    \end{algorithmic}
\end{algorithm}

\subsection{Minimum Rate Optimization}
We now consider a max-min fairness problem as follows:
\begin{equation}\label{Max_min}
\begin{split}
    \mathcal{P}_5: \;\; \underset{\mathbf{P},\mathbf{c}}{\max} \: \: & \min_{k\in\mathcal{K}}\: \: I_{\text{GMI},k}\\ 
    \text{s.t.} \: \: \: \;
    & \|\mathbf{P}\|_\text{F}^2 \leq P_\text{T}\\
    & \mathbf{c}^T \mathbf{1} = 1\\
    & \mathbf{c} \succcurlyeq \mathbf{0}.
\end{split}
\end{equation}

$\mathcal{P}_5$ is a non-convex problem. To solve it efficiently, we decouple it into the following two sub-problems,
\begin{equation}\label{Max_min_sub_c}
\begin{split}
    \mathcal{P}_6: \;\; \underset{\mathbf{c}}{\max} \: \: & \min_{k\in\mathcal{K}}\: \: I_{\text{GMI},k}(\mathbf{c};\;\mathbf{P})\\ 
    \text{s.t.} \: \: \: \;
    & \mathbf{c}^T \mathbf{1} = 1\\
    & \mathbf{c} \succcurlyeq \mathbf{0},
\end{split}
\end{equation}
and
\begin{equation}\label{Max_min_sub_P}
\begin{split}
    \mathcal{P}_7: \;\; \underset{\mathbf{P}}{\max} \: \: & \min_{k\in\mathcal{K}}\: \: I_{\text{GMI},k}(\mathbf{P};\;\mathbf{c})\\ 
    \text{s.t.} \: \: \: \;
    & \|\mathbf{P}\|_\text{F}^2 \leq P_\text{T}.
\end{split}
\end{equation}
$\mathcal{P}_6$ optimizes $\mathbf{c}$ with a fixed $\mathbf{P}$, whereas $\mathcal{P}_7$ optimizes $\mathbf{P}$ with a given $\mathbf{c}$. Note that $\mathcal{P}_6$ is a linear programming, and hence its global optimal solution can be efficiently obtained. In particular, we developed a closed-form solution as in Proposition \ref{c_opt_prop}.

\begin{proposition}\label{c_opt_prop}
    The global optimal solution of $\mathcal{P}_6$ is given in closed form by Alg. \ref{alg: c_opt}.
\end{proposition}
\begin{IEEEproof}
    See Appendix B.
\end{IEEEproof}

\begin{algorithm}[!t]
    \caption{Global Optimal Common Stream Allocation}
    \label{alg: c_opt}
    \begin{algorithmic}[1]
        \REQUIRE $I_{\text{c},k}$, $I_{\text{p},k}$, $k\in\mathcal{K}$.
        \ENSURE $\mathbf{c}^\star$
        \STATE $I_{\text{c},k}':=\min_k \;I_{\text{c},k}$ for conventional RSMA and\\ $I_{\text{c},k}':=I_{\text{c},k}$ for CS-RSMA.
        \STATE $k':=K$.
        \REPEAT 
            \STATE $\mathbf{A} := \text{diag}\{[I'_{\text{c},1},\;...,\;I'_{\text{c},k'}]\}$, $\mathbf{b} :=[I_{\text{p},1},\;...,\;I_{\text{p},k'}]^T$.
            \STATE $\xi := \frac{1+\mathbf{1}^T\mathbf{A}^{-1}\mathbf{b}}{\mathbf{1}^T\mathbf{A}^{-1}\mathbf{1}}$.
            \STATE $\mathbf{c}' := \xi\mathbf{A}^{-1}\mathbf{1} - \mathbf{A}^{-1}\mathbf{b}$.
            \STATE $k':=k'-1$.
        \UNTIL $\mathbf{c}'\succcurlyeq 0$.
        \RETURN $\mathbf{c}^\star := [\mathbf{c}'^T,\; \mathbf{0}^{1\times (K-k'-1)}]^T$.
    \end{algorithmic}
\end{algorithm}

To remove the discontinuity caused by the point-wise minimum in $\mathcal{P}_7$, we apply a log-sum-exp approximation to its objective function and consider $\mathcal{P}_8$ as follows:
\begin{equation}\label{Max_min_LSE}
\begin{split}
    \mathcal{P}_8: \;\; \underset{\mathbf{P}}{\max} \: \: & \frac{1}{\gamma}\log \sum_{k\in\mathcal{K}} \exp \: \: \left(\beta I_{\text{GMI},k}(\mathbf{P};\;\mathbf{c})\right)\\ 
    \text{s.t.} \: \: \: \;
    & \|\mathbf{P}\|_\text{F}^2 \leq P_\text{T}.
\end{split}
\end{equation}
 The gap between point-wise minimum and its log-sum-exp approximation is characterized by the following inequality \cite{Chen_2013},
 \begin{equation}
     0\leq \min_{1\leq i \leq L} f_i - \frac{1}{\gamma}\log\sum_{i=1}^L\exp(\gamma f_i) \leq \frac{1}{\gamma}\log L,\;\;\; \gamma < 0.
 \end{equation}
 
A subgradient of the objective function in $\mathcal{P}_8$ w.r.t. $\mathbf{P}$ is derived to be
\begin{equation}
\begin{split}
    g(I_\text{max-min}) = &\frac{1}{\gamma}\frac{\sum_{k\in\mathcal{K}} \beta\exp \: \: \left(\beta I_{\text{GMI},k}(\mathbf{P};\mathbf{c})\right) g(I_{\text{GMI},k}(\mathbf{P};\mathbf{c}))}{\sum_{k\in\mathcal{K}} \exp \: \: \left(\beta I_{\text{GMI},k}(\mathbf{P};\mathbf{c})\right)},
\end{split}
\end{equation}
with
\begin{equation}
    g(I_{\text{GMI},k}(\mathbf{P};\mathbf{c})) = c_k g(I_{\text{c}}) + \nabla_\mathbf{P} I_{\text{p},k},
\end{equation}
where $g(I_{\text{c}}) = \nabla_\mathbf{P} I_{\text{c},k'}^{\text{SIC/\rev{SIC-free}}}$ with $k' = \arg\min_k I_{\text{c},k'}^{\text{SIC/\rev{SIC-free}}}$ for RSMA with/without SIC, and  $g(I_{\text{c}}) = \nabla_\mathbf{P} I_{\text{c},k}^{\text{\rev{SIC-free}}}$ for CS-RSMA.

Similarly to (\ref{SR_unconstrained_form}), the inequality constraint in $\mathcal{P}_8$ can be integrated into the objective function through the logarithmic barrier. This leads to
\begin{equation}\label{Max_min_unconstrained_form}
\begin{split}
    \mathcal{P}_9: \;\; \underset{\mathbf{P}}{\max} \: \: & \frac{\tau}{\gamma}\log \sum_{k\in\mathcal{K}} \exp \: \: \left(\gamma I_{\text{GMI},k}(\mathbf{c})\right) + \log \left(P_\text{T} - \parallel\mathbf{P}\parallel_\text{F}^2\right).\\ 
\end{split}
\end{equation}
$\mathcal{P}_9$ can be solved using subgradient ascent. The solution of $\mathcal{P}_5$ is obtained by solving $\mathcal{P}_6$ and $\mathcal{P}_9$ alternatively and the steps are summarized in Alg. \ref{alg: GD for Max_min opt}, with $\Omega_\text{MMF}(\mathbf{P})$ representing the objective function in $\mathcal{P}_9$. Alg. \ref{alg: GD for Max_min opt} is guaranteed to converge because each update on $\mathbf{P}$ and $\mathbf{c}$ leads to a non-decreasing update on the objective function, and the objective function is clearly upper-bounded.

\begin{algorithm}[!t]
    \caption{Subgradient Ascent with Barrier Method for Max-Min Fairness}
    \label{alg: GD for Max_min opt}
    \begin{algorithmic}[1]
        \REQUIRE $P_\text{T}$, $\mathcal{X}_\text{c}$, $\mathcal{X}_{\text{p},k}$, $\mathbf{h}_k$, $k\in\mathcal{K}$, $\beta$, $\gamma$, $\epsilon$, $v_\text{max}$, $\tau_\text{max}$.
        \ENSURE $\mathbf{P}^\star$ and $\mathbf{c}^\star$
        \STATE Initialize $\mathbf{P}^0$ and $\tau$.
        \REPEAT 
            \STATE $v:=0$.
            \REPEAT
                \STATE Update $I_{\text{c},k}$, $I_{\text{p},k}$ based on $\mathbf{P}^v$.
                \STATE Update $\mathbf{c}$ using Alg. \ref{alg: c_opt}.
                \STATE $\Delta\mathbf{P} := \tau\:g( I_\text{max-min}) + \frac{\mathbf{P}^v}{\parallel \mathbf{P}^v \parallel_\text{F}^2 - P_\text{T}}$.
                \STATE Choose step size $\alpha$ via backtracking line search.
                \STATE Update. $\mathbf{P}^{v+1} := \mathbf{P}^v + \alpha \Delta\mathbf{P}$.
                \STATE Update. $\Omega_\text{MMF}^{v+1} := \Omega_\text{MMF}(\mathbf{P}^{v+1})$.
                \STATE $v:=v+1$.
            \UNTIL $\mid \Omega_\text{MMF}^{v} - \Omega_\text{MMF}^{,v-1} \mid\; < \epsilon$ or $v \geq v_\text{max}$.
            \STATE Update $\tau := \beta \tau$.
        \UNTIL $\tau \geq \tau_\text{max}$.
        \RETURN $\mathbf{P}^\star := \mathbf{P}^{v}$ and $\mathbf{c}^\star := \mathbf{c}$.
    \end{algorithmic}
\end{algorithm}
%---------------------------------------------------%
\section{Numerical Results on\\ Precoder Optimization for RSMA}\label{Sec: Opt results}
In this section, we investigate the theoretical performance of CS-RSMA and conventional RSMA schemes under mismatched decoding through numerical simulations. Prior to the results, adaptive modulation and channel model will be introduced.

All the results in this section are ergodic performance measured by averaging 100 channel realizations. The precoder initialization in Alg. \ref{alg: GD for SR opt} and \ref{alg: GD for Max_min opt} involves the low-complexity precoders proposed in \cite{Sibo_TCOM} for constellation-constrained sum-rate optimization and some randomly generated precoders.

\subsection{Adaptive modulation}
We assume that all the users are subject to a decoding complexity $\delta=4$ or $16$.\footnote{It is assumed that $\delta$ is the same for all the users for simplicity and clarity. The problem and solution introduced in Section \ref{sec: Opt algo} can be trivially generalized to ones with non-uniform complexity constraints among users.} The alphabets applied to the streams are chosen from BPSK and QAM constellations, subject to the decoding complexity constraints. This leads to a few transmission modes in terms of alphabets for each complexity constraint, as summarized in Tables \ref{Table: Tx mode n=4} and \ref{Table: Tx mode n=16}. The simulations allow adaptive alphabet selection under certain decoding complexity.

\begin{remark}
    With Mode 1 in both Tables \ref{Table: Tx mode n=4} and \ref{Table: Tx mode n=16}, RSMA becomes SDMA; with Mode 3 in Table \ref{Table: Tx mode n=4} and Mode 5 in Table \ref{Table: Tx mode n=16}, RSMA is equivalent to multicasting all messages \cite{Clerckx_WCL}.
\end{remark}

\begin{remark}\label{Remark: upper bound on SR}
    Achievable rates under finite alphabets are upper-limited by the cardinality of the alphabet. Different transmission modes lead to different maximum achievable rates. It can be observed from Tables \ref{Table: Tx mode n=4} and \ref{Table: Tx mode n=16} that the sum rate of Mode 1, i.e., SDMA, is upper-bounded by $K \log_2 \delta$ bits and is the highest among all the modes, whereas the sum rate of multicasting, i.e., the last mode in both tables, is upper-bounded by $\log_2 \delta$ bits and is the lowest among all the modes. For example, for $K=2$ and $\delta=16$, Mode 1 can bring at most 8 bits per transmission in SR, which any other mode cannot achieve.
\end{remark}

\begin{table}
    \centering
    \caption{Transmission modes under decoding complexity $\delta=4$}
    \begin{tabular}{|c|c|c|c|c|}
        \hline
        Mode & 1 & 2 & 3 \\
        \hline
        Private Stream &  QPSK & BPSK & $\{0\}$\\
        \hline
        Common Stream & $\{0\}$ & BPSK & QPSK \\
        \hline
    \end{tabular}
    \label{Table: Tx mode n=4}
\end{table}

\begin{table}
    \centering
    \caption{Transmission modes under decoding complexity $\delta=16$}
    \begin{tabular}{|c|c|c|c|c|c|c|}
        \hline
        Mode & 1 & 2 & 3 & 4 & 5\\
        \hline
        Private Stream &  16QAM & 8QAM & QPSK & BPSK & \{0\}\\
        \hline
        Common Stream & $\{0\}$ & BPSK & QPSK & 8QAM & 16QAM\\
        \hline
    \end{tabular}
    \label{Table: Tx mode n=16}
\end{table}

\subsection{Channel model}
The spatially correlated Rayleigh-fading channel is given by $\mathbf{h}_k \sim \mathcal{CN}(\mathbf{0}^{N_\text{T} \times 1},\;\mathbf{R}_k)$, where $\mathbf{R}_k \in \mathbb{C}^{N_\text{T} \times N_\text{T}}$ is the channel covariance matrix for user-$k$. With eigendecomposision to $\mathbf{R}_k$, 
\begin{equation}
    \mathbf{R}_k = \mathbf{U}_k \mathbf{\Lambda}_k \mathbf{U}_k^H, 
\end{equation}
where $\mathbf{\Lambda}_k$ is a $r_k \times r_k$ diagonal matrix containing the nonzero eigenvalues of $\mathbf{R}_k$ and $\mathbf{U}_k \in \mathbb{C}^{N_\text{t} \times r_k} $ contains the eigenvectors of $\mathbf{R}_k$ corresponding to the nonzero eigenvalues. The Karhunen-Loeve representation of $\mathbf{h}_k$ gives
\begin{equation}
    \mathbf{h}_k = \mathbf{U}_k\mathbf{\Lambda}_k^{\frac{1}{2}}\mathbf{w}_k,
\end{equation}
where $\mathbf{w}_k \in \mathbb{C}^{r_k \times 1} \sim \mathcal{CN}(\mathbf{0}^{r_k \times 1},\;\mathbf{I}^{r_k})$.

We follow the one-ring scattering model proposed in \cite{one_ring_model}, which assumes that there is no line-of-sight path between the transmitter and user-$k$ and the angle-of-departure of all scatters is uniformly distributed around a centre angle-of-departure, $\theta_k$, with an angular spread of $\Delta_k$ for user-$k$. Consider a uniform and linear antenna array with the spacing between the antenna elements to be $\lambda/2$, where $\lambda$ denotes the wavelength. The elements of $\mathbf{R}_k$ are given by
\begin{equation}
    [\mathbf{R}_k]_{m,n} = \frac{1}{2\Delta_k} \int_{-\Delta_k}^{\Delta_k}
    e^{-j\pi (\alpha+\theta_k)(m-n) \sin{\alpha}} d\alpha.
\end{equation}
We assume that $\theta_k=\pi/3$ and $\Delta_k=\Delta$, $\forall k\in\mathcal{K}$, which implies that users with similar channel statistics are served at the same time. The purpose of this setting is to model communication scenarios with a certain user density.

\subsection{Sum-rate evaluation}

\begin{figure}
    \centering
    \subfigure[$N_\text{T}=2$, $K=2$, $\Delta=\pi/18$, $\delta=16$.]{\includegraphics[width=4.2cm]{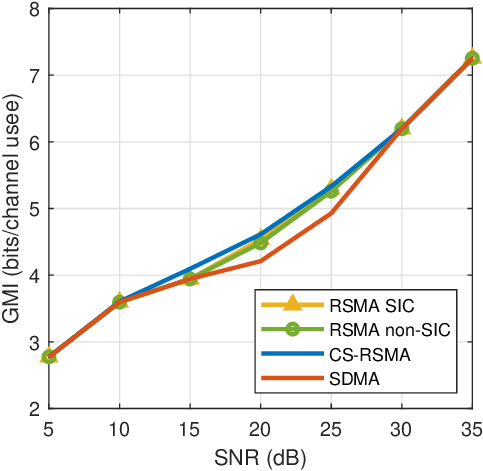}}\hspace{0cm}
    \subfigure[$N_\text{T}=4$, $K=4$, $\Delta=\pi/12$, $\delta=4$.]{\includegraphics[width=4.2cm]{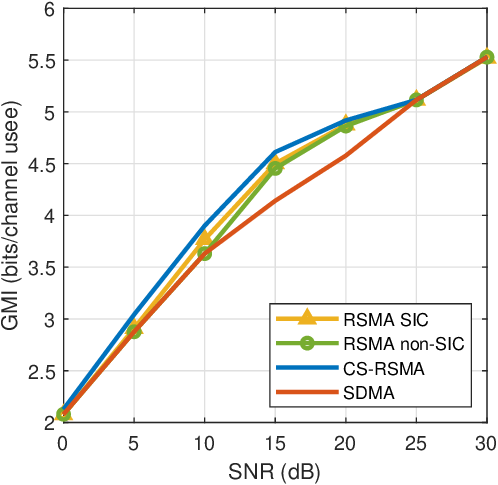}}\hspace{0cm}
    \caption{Ergodic SR performance of conventional RSMA, CS-RSMA and SDMA.}
    \label{Fig: SR scheme comparision}

    \centering
    \subfigure[$N_\text{T}=2$, $K=2$, $\Delta=\pi/18$, $\delta=16$.]{\includegraphics[width=4.2cm]{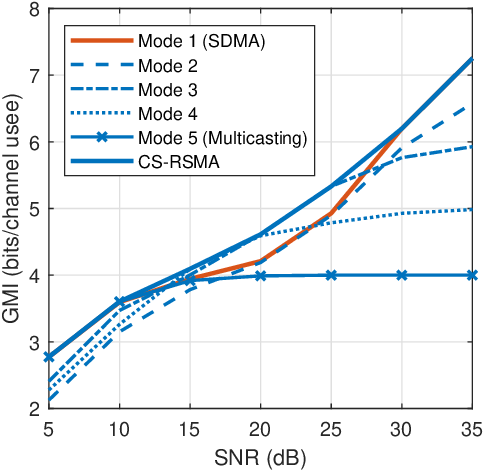}}\hspace{0cm}
    \subfigure[$N_\text{T}=4$, $K=4$, $\Delta=\pi/12$, $\delta=4$.]{\includegraphics[width=4.2cm]{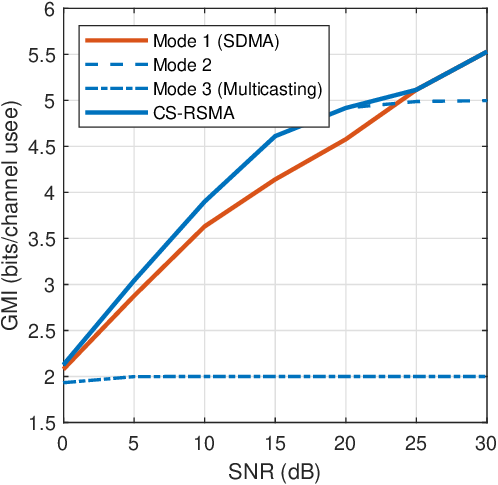}}\hspace{0cm}
    \caption{Ergodic SR performance of different modes of CS-RSMA.}
    \label{Fig: SR mode comparision}
\end{figure}

\begin{figure}
    \centering
    \includegraphics[width=7.2cm]{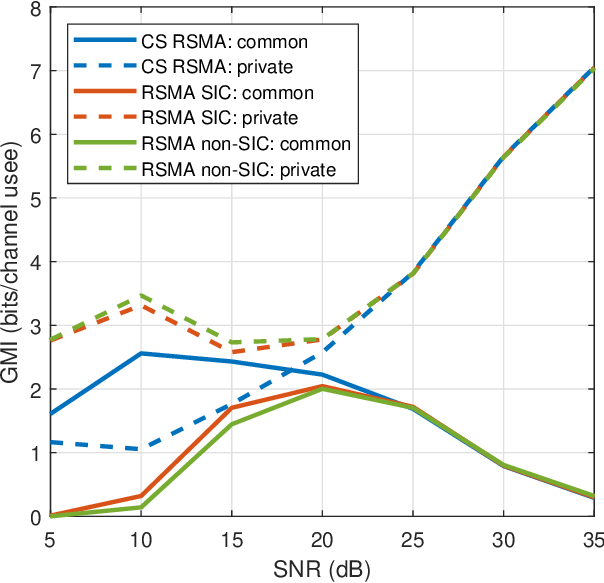}
    \caption{\rev{Comparing contribution of common and private stream in achievable rate} with $N_\text{T}=2$, $K=2$, $\Delta=\pi/18$, $\delta=16$.}
    \label{Fig: SR stream comparision}
\end{figure}

Fig. \ref{Fig: SR scheme comparision} depicts the ergodic SR performance of conventional RSMA schemes, CS-RSMA and SDMA. It can be observed that RSMA schemes, including CS-RSMA, bring benefits in the medium SNR regime compared to SDMA. Furthermore, CS-RSMA slightly outperforms conventional RSMA schemes. The ergodic SR led by different transmission modes under CS-RSMA is depicted in Fig. \ref{Fig: SR mode comparision}, together with the performance of CS-RSMA given by adaptive alphabet/mode selection as a reference. It can be observed that the transmission modes that enable the common stream achieve higher SR than Mode 1, i.e., SDMA, at medium SNR. At high SNR, Mode 1 dominates as other modes saturate at lower SR, as suggested by Remark \ref{Remark: upper bound on SR}.

Fig. \ref{Fig: SR stream comparision} depicts the ergodic rates of common and private streams under CS-RSMA and conventional RSMA. It can be observed that CS-RSMA tends to carry higher information rates in the common stream than conventional RSMA. This is a consequence of having ``$\max$" on the common rates in the SR problem formulation for CS-RSMA instead of the ``$\min$" in the one for conventional RSMA, as shown in $\mathcal{P}_2$ and $\mathcal{P}_3$.

%Interesting discussion but need to see how to enable it.
%Intuitively, the reason that activating common stream at medium SNR is as follows. From low SNR to medium SNR, add more power to the private stream is no longer beneficial because that private stream is saturating due to the support of the alphabet. Allocating high power to the second private streams brings inter-user interference which limits the overall performance. Such interference is even more harmful when they are considered to be Gaussian distributed by users. On the other hand, the common stream is treated optimally by the decoders and hence brings is less harmful to the decoding of the private streams. Hence, with limited SNR, activating the common stream is more beneficial to the overall performance. 

\subsection{Max-min fairness optimization}

\begin{figure}
    \centering
    \subfigure[$N_\text{T}=2$, $K=2$, $\Delta=\pi/18$, $\delta=16$.]{\includegraphics[width=4.35cm]{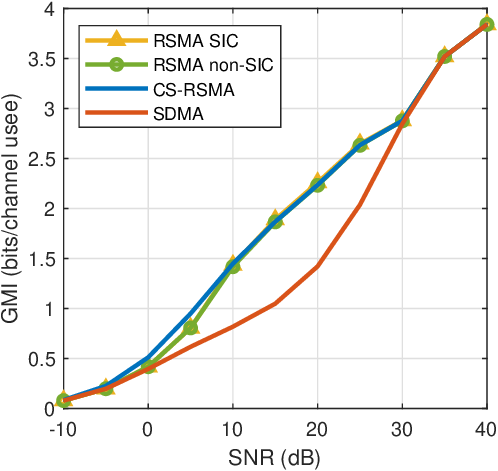}}\hspace{0cm}
    \subfigure[$N_\text{T}=4$, $K=4$, $\Delta=\pi/9$, $\delta=4$.]{\includegraphics[width=4.35cm]{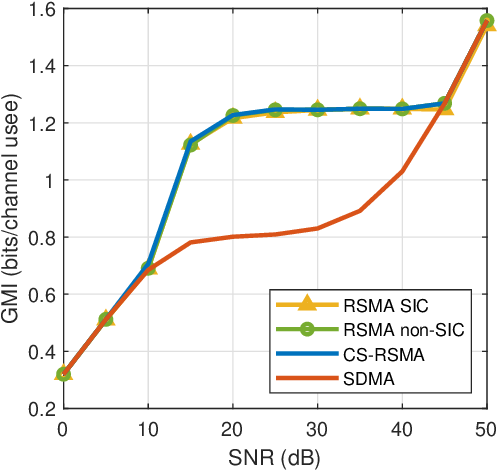}}\hspace{0cm}
    \caption{Ergodic MMF performance of conventional RSMA, CS-RSMA and SDMA.}
    \label{Fig: MMF scheme comparision}

    \centering
    \subfigure[$N_\text{T}=2$, $K=2$, $\Delta=\pi/18$, $\delta=16$.]{\includegraphics[width=4.35cm]{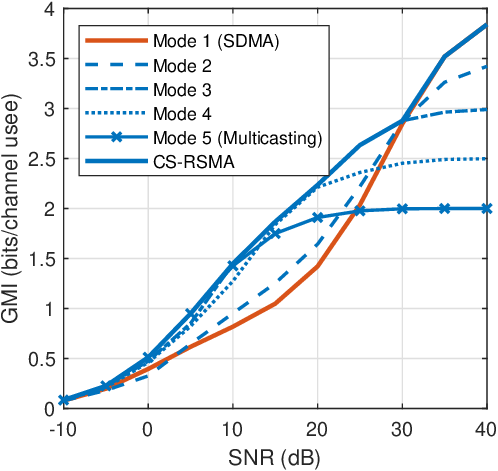}}\hspace{0cm}
    \subfigure[$N_\text{T}=4$, $K=4$, $\Delta=\pi/9$, $\delta=4$.]{\includegraphics[width=4.35cm]{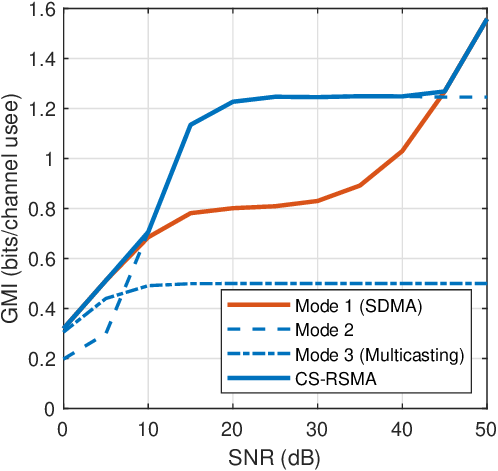}}\hspace{0cm}
    \caption{Ergodic MMF performance of different modes of CS-RSMA.}
    \label{Fig: MMF mode comparision}
\end{figure}

Fig. \ref{Fig: MMF scheme comparision} depicts the ergodic MMF performance of conventional RSMA schemes, CS-RSMA and SDMA. Similar to SR performance, RSMA brings benefits at the medium SNR regime by activating the common stream, but converges to SDMA at the high SNR regime. CS-RSMA performs very similar to conventional RSMA schemes.\footnote{For both SR and MMF evaluation, if the input signals are not from finite alphabets but are Gaussian distributed, the gap between RSMA with SIC and SDMA exists in both medium and high SNR regime, as shown in \cite{Hamdi,Hamdi_MMF}, because \rev{the instantaneous information rate} of the common stream will not saturate at high SNR. CS-RSMA, \rev{SIC-free} RSMA and SDMA will perform similarly under Gaussian input, because the interference from the common stream when decoding the private stream is more detrimental than with finite-alphabet inputs.}  The performance of different modes of CS-RSMA is depicted in Fig. \ref{Fig: MMF mode comparision}, along with the performance of CS-RSMA given by adaptive alphabet/mode selection as a reference. It shows that activating the common stream brings higher minimum rates among users under medium SNR than Mode 1, i.e., SDMA, whereas Mode 1 dominates under high SNR, because the modes enabling the common stream saturates at lower rates, as suggested by Remark \ref{Remark: upper bound on SR}.

%It can be observed that SNR regime where RSMA brings benefit over SDMA enlarges in comparison to the sum-rate results, especially at low SNR. This is because, at low SNR, activating only one private stream is does not meet the fairness requirement. Activating both private streams would cause inter-user interference to the two users and hence harms the performance. Activating the common stream is more beneficial as it is less harmful than private streams.

\section{PHY Transceiver Design for CS-RSMA\\and Link-Level Simulation}\label{Sec: PHY TxRx and LLS}
In this section, we propose a transceiver architecture for CS-RSMA and evaluate its performance using link-level simulations.

\subsection{PHY Transceiver Design}
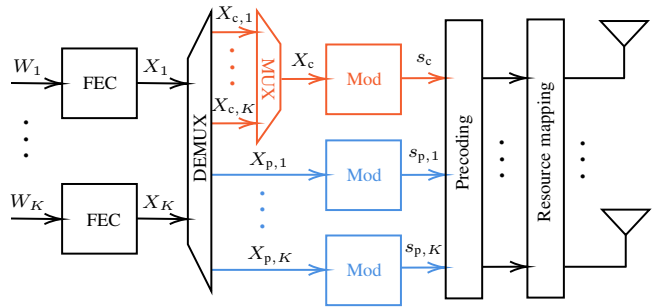
\begin{figure}
\centering

\tikzset{every picture/.style={line width=0.75pt}} %set default line width to 0.75pt        

\begin{tikzpicture}[x=0.6pt,y=0.6pt,yscale=-1,xscale=1]

%Straight Lines [id:da05949504436426367] 
\draw[->] [color={rgb, 255:red, 74; green, 144; blue, 226 }  ,draw opacity=1 ][line width=0.75]    (132,200) -- (204.18,200) ;

%Shape: Rectangle [id:dp5144059547225027] 
\draw  [line width=0.75]  (37.68,60.31) -- (84.78,60.31) -- (84.78,104.26) -- (37.68,104.26) -- cycle ;
%Straight Lines [id:da22094917823176496] 
\draw[->] [line width=0.75]    (84.78,82.09) -- (116.95,82.46) ;

%Straight Lines [id:da460683455088177] 
\draw[->] [line width=0.75]    (5.46,82.46) -- (37.68,82.46) ;

%Shape: Rectangle [id:dp4149872738079208] 
\draw  [line width=0.75]  (37.68,145.81) -- (84.78,145.81) -- (84.78,189.76) -- (37.68,189.76) -- cycle ;
%Straight Lines [id:da9591458116804961] 
\draw[->] [line width=0.75]    (84.46,167.96) -- (116.95,167.96) ;

%Straight Lines [id:da7209203425891032] 
\draw[->] [line width=0.75]    (5.46,167.59) -- (37.68,167.96) ;

%Shape: Rectangle [id:dp5870742923493278] 
\draw  [color={rgb, 255:red, 243; green, 93; blue, 45 }  ,draw opacity=1 ][line width=0.75]  (204.18,57.81) -- (251.28,57.81) -- (251.28,101.76) -- (204.18,101.76) -- cycle ;
%Shape: Trapezoid [id:dp8247038909961834] 
\draw  [color={rgb, 255:red, 243; green, 93; blue, 45 }  ,draw opacity=1 ][line width=0.75]  (160.5,36.5) -- (175.31,57.5) -- (175.31,99) -- (160.5,120) -- cycle ;
%Straight Lines [id:da40580780862291477] 
\draw[->] [color={rgb, 255:red, 243; green, 93; blue, 45 }  ,draw opacity=1 ][line width=0.75]    (131.96,50.09) -- (160.5,50.01) ;

%Straight Lines [id:da8624369234080465] 
\draw[->] [color={rgb, 255:red, 243; green, 93; blue, 45 }  ,draw opacity=1 ][line width=0.75]    (131.96,107.59) -- (160.5,107.51) ;

%Straight Lines [id:da04072679190311612] 
\draw[->] [color={rgb, 255:red, 243; green, 93; blue, 45 }  ,draw opacity=1 ][line width=0.75]    (175.96,79.59) -- (204.18,79.51) ;

%Shape: Rectangle [id:dp29139493188861276] 
\draw  [color={rgb, 255:red, 74; green, 144; blue, 226 }  ,draw opacity=1 ][line width=0.75]  (204.18,118.31) -- (251.28,118.31) -- (251.28,162.26) -- (204.18,162.26) -- cycle ;
%Straight Lines [id:da60624236029415] 
\draw[->] [color={rgb, 255:red, 74; green, 144; blue, 226 }  ,draw opacity=1 ][line width=0.75]    (132,140) -- (204.18,140) ;

%Shape: Rectangle [id:dp8281833706374933] 
\draw  [color={rgb, 255:red, 74; green, 144; blue, 226 }  ,draw opacity=1 ][line width=0.75]  (204.18,178.31) -- (251.28,178.31) -- (251.28,222.26) -- (204.18,222.26) -- cycle ;
%Shape: Trapezoid [id:dp170488783878063] 
\draw  [line width=0.75]  (131.77,213.5) -- (116.95,183.87) -- (116.95,66.64) -- (131.77,37) -- cycle ;
%Straight Lines [id:da18206471093930343] 
\draw[->] [color={rgb, 255:red, 243; green, 93; blue, 45 }  ,draw opacity=1 ][line width=0.75]    (251.46,79.09) -- (279.5,79.01) ;

%Straight Lines [id:da5878884052143541] 
\draw[->] [color={rgb, 255:red, 74; green, 144; blue, 226 }  ,draw opacity=1 ][line width=0.75]    (251.46,140.09) -- (279.5,140.01) ;

%Straight Lines [id:da4755403659326355] 
\draw[->] [color={rgb, 255:red, 74; green, 144; blue, 226 }  ,draw opacity=1 ][line width=0.75]    (251.46,198.59) -- (279.5,198.51) ;

%Shape: Rectangle [id:dp7879623237471022] 
\draw   (279.5,218) -- (279.5,42.5) -- (302.5,42.5) -- (302.5,218) -- cycle ;
%Straight Lines [id:da8575360436163839] 
\draw[->] [color={rgb, 255:red, 0; green, 0; blue, 0 }  ,draw opacity=1 ][line width=0.75]    (302.46,79.09) -- (331,79.01) ;

%Straight Lines [id:da3008445168301772] 
\draw[->] [color={rgb, 255:red, 0; green, 0; blue, 0 }  ,draw opacity=1 ][line width=0.75]    (302.96,198.09) -- (331,198.01) ;

%Shape: Rectangle [id:dp9176257871570053] 
\draw   (331,217.5) -- (331,42) -- (354,42) -- (354,217.5) -- cycle ;
%Shape: Triangle [id:dp8707096485725488] 
\draw  [line width=0.75]  (392,59) -- (377.1,43) -- (406.9,43) -- cycle ;
%Straight Lines [id:da038508025024558634] 
\draw [line width=0.75]    (392,59) -- (392,79) ;
%Straight Lines [id:da6133350893990279] 
\draw [line width=0.75]    (353.48,79.04) -- (392.48,79.04) ;
%Shape: Triangle [id:dp11094258140626978] 
\draw  [line width=0.75]  (393,178) -- (378.1,162) -- (407.9,162) -- cycle ;
%Straight Lines [id:da5129233911687984] 
\draw [line width=0.75]    (393,178) -- (393,198) ;
%Straight Lines [id:da037248204755312875] 
\draw [line width=0.75]    (354.48,198.04) -- (393.48,198.04) ;

% Text Node
\draw (61.67,82.13) node  [font=\scriptsize] [align=left] {\begin{minipage}[lt]{17pt}\setlength\topsep{0pt}
\begin{center}
FEC
\end{center}

\end{minipage}};
% Text Node
\draw (16.18,71.95) node  [font=\scriptsize] [align=left] {\begin{minipage}[lt]{13.91pt}\setlength\topsep{0pt}
\begin{center}
$\displaystyle W_{1}$
\end{center}

\end{minipage}};
% Text Node
\draw (61.67,168.13) node  [font=\scriptsize] [align=left] {\begin{minipage}[lt]{19.42pt}\setlength\topsep{0pt}
\begin{center}
FEC
\end{center}

\end{minipage}};
% Text Node
\draw (16.18,157.45) node  [font=\scriptsize] [align=left] {\begin{minipage}[lt]{14.85pt}\setlength\topsep{0pt}
\begin{center}
$\displaystyle W_{K}$
\end{center}

\end{minipage}};
% Text Node
\draw (97,72) node  [font=\scriptsize] [align=left] {\begin{minipage}[lt]{12.11pt}\setlength\topsep{0pt}
\begin{center}
$\displaystyle X_{1}$
\end{center}

\end{minipage}};
% Text Node
\draw (99,155) node  [font=\scriptsize] [align=left] {\begin{minipage}[lt]{13.06pt}\setlength\topsep{0pt}
\begin{center}
$\displaystyle X_{K}$
\end{center}

\end{minipage}};
% Text Node
\draw (15,117.48) node  [font=\Large,rotate=-269.91] [align=left] {. . .};
% Text Node
\draw (124.66,123.01) node  [font=\scriptsize,rotate=-269.99] [align=left] {DEMUX};
% Text Node
\draw (228.17,79.63) node  [font=\scriptsize,color={rgb, 255:red, 243; green, 93; blue, 45 }  ,opacity=1 ] [align=left] {\begin{minipage}[lt]{16.62pt}\setlength\topsep{0pt}
\begin{center}
Mod
\end{center}

\end{minipage}};
% Text Node
\draw (173.5,64.14) node [anchor=north west][inner sep=0.75pt]  [font=\scriptsize,color={rgb, 255:red, 243; green, 93; blue, 45 }  ,opacity=1 ,rotate=-90] [align=left] {MUX};
% Text Node
\draw (146.68,39.45) node  [font=\scriptsize] [align=left] {\begin{minipage}[lt]{16.02pt}\setlength\topsep{0pt}
\begin{center}
$\displaystyle X_{\text{c} ,1}$
\end{center}

\end{minipage}};
% Text Node
\draw (145.68,94.5) node  [font=\scriptsize] [align=left] {\begin{minipage}[lt]{16.97pt}\setlength\topsep{0pt}
\begin{center}
$\displaystyle X_{\text{c} ,K}$
\end{center}

\end{minipage}};
% Text Node
\draw (145,72) node  [font=\Large,color={rgb, 255:red, 243; green, 93; blue, 45 }  ,opacity=1 ,rotate=-269.91] [align=left] {. . .};
% Text Node
\draw (189.68,68) node  [font=\scriptsize] [align=left] {\begin{minipage}[lt]{11.78pt}\setlength\topsep{0pt}
\begin{center}
$\displaystyle X_{\text{c}}$
\end{center}

\end{minipage}};
% Text Node
\draw (228.17,140.13) node  [font=\scriptsize,color={rgb, 255:red, 74; green, 144; blue, 226 }  ,opacity=1 ] [align=left] {\begin{minipage}[lt]{16.62pt}\setlength\topsep{0pt}
\begin{center}
Mod
\end{center}

\end{minipage}};
% Text Node
\draw (228.17,200.13) node  [font=\scriptsize,color={rgb, 255:red, 74; green, 144; blue, 226 }  ,opacity=1 ] [align=left] {\begin{minipage}[lt]{16.62pt}\setlength\topsep{0pt}
\begin{center}
Mod
\end{center}

\end{minipage}};
% Text Node
\draw (168.68,188.5) node  [font=\scriptsize] [align=left] {\begin{minipage}[lt]{17.24pt}\setlength\topsep{0pt}
\begin{center}
$\displaystyle X_{\text{p} ,K}$
\end{center}

\end{minipage}};
% Text Node
\draw (168.18,131.45) node  [font=\scriptsize] [align=left] {\begin{minipage}[lt]{16.3pt}\setlength\topsep{0pt}
\begin{center}
$\displaystyle X_{\text{p} ,1}$
\end{center}

\end{minipage}};
% Text Node
\draw (164,162) node  [font=\Large,rotate=-269.91] [align=left] {\textcolor[rgb]{0.29,0.56,0.89}{. . .}};
% Text Node
\draw (266.18,67.45) node  [font=\scriptsize] [align=left] {\begin{minipage}[lt]{9.08pt}\setlength\topsep{0pt}
\begin{center}
$\displaystyle s_{\text{c}}$
\end{center}

\end{minipage}};
% Text Node
\draw (266.18,127.95) node  [font=\scriptsize] [align=left] {\begin{minipage}[lt]{13.6pt}\setlength\topsep{0pt}
\begin{center}
$\displaystyle s_{\text{p} ,1}$
\end{center}

\end{minipage}};
% Text Node
\draw (265.68,183.5) node  [font=\scriptsize] [align=left] {\begin{minipage}[lt]{14.55pt}\setlength\topsep{0pt}
\begin{center}
$\displaystyle s_{\text{p} ,K}$
\end{center}

\end{minipage}};
% Text Node
\draw (291.16,126.01) node  [font=\scriptsize,rotate=-269.99] [align=left] {Precoding};
% Text Node
\draw (315,129.98) node  [font=\Large,rotate=-269.91] [align=left] {. . .};
% Text Node
\draw (342.66,125.51) node  [font=\scriptsize,rotate=-269.99] [align=left] {Resource mapping};
% Text Node
\draw (366,130.48) node  [font=\Large,rotate=-269.91] [align=left] {. . .};
\end{tikzpicture}
\caption{PHY transmitter design for CS-RSMA.}
\label{PHY TX}
\end{figure}

Fig. \ref{PHY TX} depicts the proposed CS-RSMA transmitter architecture, where ``FEC" and ``Mod" represent forward error correction coding\footnote{This potentially includes interleaving and scrambling.} and modulation (mapping), respectively. For clarity, the time index is neglected in the notation. In Fig. \ref{PHY TX}, the messages, $W_{1}$, $W_{2}$,\;..., $W_{K}$, are first encoded into binary codewords, denoted by $X_{1}$, $X_{2}$,\;..., $X_{K}$. Codeword segmentation and combining can be achieved by de-multiplexing and multiplexing the encoded bits into $X_{\text{c}}$, $X_{\text{p},1}$, $X_{\text{p},2}$,..., $X_{\text{p},K}$. The common and private streams are obtained by modulating the resulting $K+1$ bit streams into symbol streams. The precoders then mix the symbol streams into a stream of vector symbols to feed the multiple transmit antennas. The resource mapper allocates different instances of the vector symbol stream to different physical layer resource elements\footnote{These can be created, for example, by Orthogonal Frequency-Division Multiplexing (OFDM) and Time Domain Multiplexing (TDM).}.
%\todo{May be change all the notation for sequence of bits into bold letter? Check section II.}

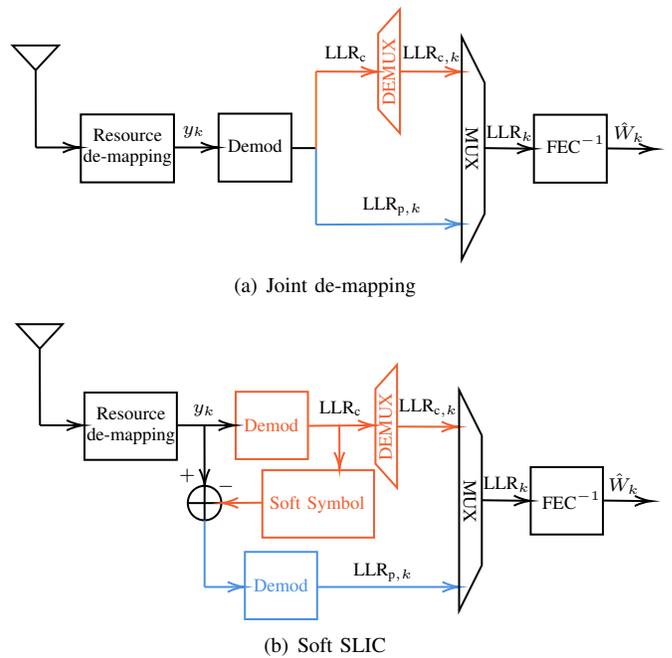
\begin{figure}
\centering
\subfigure[Joint de-mapping]{

\tikzset{every picture/.style={line width=0.75pt}} %set default line width to 0.75pt        

\begin{tikzpicture}[x=0.75pt,y=0.75pt,yscale=-0.78,xscale=0.78]

%Shape: Triangle [id:dp8517202042864013] 
\draw  [line width=0.75]  (22,93.5) -- (7.1,77.5) -- (36.9,77.5) -- cycle ;
%Straight Lines [id:da6933552268686084] 
\draw [line width=0.75]    (22.15,93.5) -- (22.15,143.49) ;
%Straight Lines [id:da5531028034439657] 
\draw[->] [line width=0.75]    (22.15,143.49) -- (51.19,143.49) ;

%Shape: Rectangle [id:dp9714204165927226] 
\draw  [line width=0.75]  (51.19,122.11) -- (111.43,122.11) -- (111.43,166.06) -- (51.19,166.06) -- cycle ;
%Shape: Rectangle [id:dp15004146806773566] 
\draw  [line width=0.75]  (140.29,121.61) -- (187.4,121.61) -- (187.4,165.56) -- (140.29,165.56) -- cycle ;
%Straight Lines [id:da2598549447809232] 
\draw [line width=0.75]    (187.9,143.89) -- (203.28,143.89) ;
%Straight Lines [id:da48415553769771413] 
\draw [color={rgb, 255:red, 74; green, 144; blue, 226 }  ,draw opacity=1 ][line width=0.75]    (203.28,193.42) -- (203.28,143.89) ;
%Straight Lines [id:da8469345004605622] 
\draw[->] [color={rgb, 255:red, 74; green, 144; blue, 226 }  ,draw opacity=1 ][line width=0.75]    (203.28,193.42) -- (297.5,193.01) ;

%Shape: Trapezoid [id:dp15488443489939596] 
\draw  [color={rgb, 255:red, 243; green, 93; blue, 45 }  ,draw opacity=1 ][line width=0.75]  (257.55,134.31) -- (243.45,118.6) -- (243.45,69.71) -- (257.55,54) -- cycle ;
%Straight Lines [id:da8396375751214609] 
\draw [color={rgb, 255:red, 245; green, 108; blue, 35 }  ,draw opacity=1 ][fill={rgb, 255:red, 245; green, 166; blue, 35 }  ,fill opacity=1 ][line width=0.75]    (203.28,143.89) -- (203.28,94.35) ;
%Straight Lines [id:da4943871418722343] 
\draw[->] [color={rgb, 255:red, 243; green, 93; blue, 45 }  ,draw opacity=1 ][line width=0.75]    (257.55,94.35) -- (297.5,94.49) ;

%Straight Lines [id:da9511236534767346] 
\draw[->] [line width=0.75]    (312.13,143.89) -- (344.18,144.26) ;

%Shape: Rectangle [id:dp8719426310322989] 
\draw  [line width=0.75]  (344.18,122.31) -- (391.28,122.31) -- (391.28,166.26) -- (344.18,166.26) -- cycle ;
%Straight Lines [id:da21680664599894006] 
\draw[->] [line width=0.75]    (390.96,144.09) -- (421,144.46) ;

%Shape: Trapezoid [id:dp042112864675418527] 
\draw  [line width=0.75]  (297.5,72) -- (312.31,101.63) -- (312.31,186.37) -- (297.5,216) -- cycle ;
%Straight Lines [id:da9952029720711312] 
\draw[->] [line width=0.75]    (111.32,143.49) -- (140.29,143.5) ;

%Straight Lines [id:da8938716734221143] 
\draw[->] [color={rgb, 255:red, 243; green, 93; blue, 45 }  ,draw opacity=1 ][line width=0.75]    (203.28,94.35) -- (243.45,94.49) ;

% Text Node
\draw (45,129.5) node [anchor=north west][inner sep=0.75pt]  [font=\scriptsize] [align=left] {\begin{minipage}[lt]{40.43pt}\setlength\topsep{0pt}
\begin{center}
Resource\\de-mapping
\end{center}

\end{minipage}};
% Text Node
\draw (140,136) node [anchor=north west][inner sep=0.75pt]  [font=\scriptsize] [align=left] {\begin{minipage}[lt]{25.74pt}\setlength\topsep{0pt}
\begin{center}
Demod
\end{center}

\end{minipage}};
% Text Node
\draw (207.26,76) node [anchor=north west][inner sep=0.75pt]  [font=\scriptsize] [align=left] {LLR$\displaystyle _{\text{c}}$};
% Text Node
\draw (257.57,76) node [anchor=north west][inner sep=0.75pt]  [font=\scriptsize] [align=left] {LLR$\displaystyle _{\text{c} ,k}$};
% Text Node
\draw (231.68,174) node [anchor=north west][inner sep=0.75pt]  [font=\scriptsize] [align=left] {LLR$\displaystyle _{\text{p} ,k}$};
% Text Node
\draw (244.5,118.35) node [anchor=north west][inner sep=0.75pt]  [font=\scriptsize,color={rgb, 255:red, 243; green, 93; blue, 45 } ,rotate=-270] [align=left] {DEMUX};
% Text Node
\draw (310,128.5) node [anchor=north west][inner sep=0.75pt]  [font=\scriptsize,rotate=-90] [align=left] {MUX};
% Text Node
\draw (348.67,133.5) node [anchor=north west][inner sep=0.75pt]  [font=\scriptsize] [align=left] {\begin{minipage}[lt]{24.19pt}\setlength\topsep{0pt}
\begin{center}
FEC$\displaystyle ^{-1}$
\end{center}

\end{minipage}};
% Text Node
\draw (391.95,125) node [anchor=north west][inner sep=0.75pt]  [font=\scriptsize] [align=left] {\begin{minipage}[lt]{13.73pt}\setlength\topsep{0pt}
\begin{center}
$\displaystyle \hat{W}_{k}$
\end{center}

\end{minipage}};
% Text Node
\draw (311.77,127.5) node [anchor=north west][inner sep=0.75pt]  [font=\scriptsize] [align=left] {LLR$\displaystyle _{k}$};
% Text Node
\draw (115.13,127.5) node [anchor=north west][inner sep=0.75pt]  [font=\scriptsize] [align=left] {$\displaystyle y_{k}$};

\end{tikzpicture}}

\subfigure[Soft SLIC]{

\tikzset{every picture/.style={line width=0.75pt}} %set default line width to 0.75pt        

\begin{tikzpicture}[x=0.58pt,y=0.58pt,yscale=-1,xscale=1]

%Shape: Triangle [id:dp8517202042864013] 
\draw  [line width=0.75]  (16,93.5) -- (1.1,77.5) -- (30.9,77.5) -- cycle ;
%Straight Lines [id:da6933552268686084] 
\draw [line width=0.75]    (16,93.5) -- (16.15,143.49) ;
%Straight Lines [id:da5531028034439657] 
\draw [line width=0.75][->]    (16.15,143.49) -- (45.19,143.5) ;

%Shape: Rectangle [id:dp9714204165927226] 
\draw  [line width=0.75]  (45.19,122.11) -- (105.43,122.11) -- (105.43,166.06) -- (45.19,166.06) -- cycle ;
%Shape: Rectangle [id:dp15004146806773566] 
\draw  [color={rgb, 255:red, 243; green, 93; blue, 45 }  ,draw opacity=1 ][line width=0.75]  (143.79,121.61) -- (190.9,121.61) -- (190.9,165.56) -- (143.79,165.56) -- cycle ;
%Straight Lines [id:da8469345004605622] 
\draw[->] [color={rgb, 255:red, 74; green, 144; blue, 226 }  ,draw opacity=1 ][line width=0.75]    (196.5,249) -- (289.5,249) ;

%Shape: Trapezoid [id:dp15488443489939596] 
\draw  [color={rgb, 255:red, 243; green, 93; blue, 45 }  ,draw opacity=1 ][line width=0.75]  (249.05,184.31) -- (234.95,168.6) -- (234.95,119.71) -- (249.05,104) -- cycle ;
%Straight Lines [id:da4943871418722343] 
\draw [->] [color={rgb, 255:red, 243; green, 93; blue, 45 }  ,draw opacity=1 ][line width=0.75]    (249.05,144.35) -- (289.5,144.49) ;

%Straight Lines [id:da9511236534767346] 
\draw[->] [line width=0.75]    (304.13,192.39) -- (336.18,192.76) ;

%Shape: Rectangle [id:dp8719426310322989] 
\draw  [line width=0.75]  (336.18,170.81) -- (383.28,170.81) -- (383.28,214.76) -- (336.18,214.76) -- cycle ;
%Straight Lines [id:da21680664599894006] 
\draw[->] [line width=0.75]    (382.96,192.59) -- (413,192.96) ;

%Shape: Trapezoid [id:dp042112864675418527] 
\draw  [line width=0.75]  (289.5,120.5) -- (304.31,150.13) -- (304.31,234.87) -- (289.5,264.5) -- cycle ;
%Straight Lines [id:da9952029720711312] 
\draw[->] [line width=0.75]    (105.5,143.5) -- (143.79,143.5) ;

%Straight Lines [id:da8938716734221143] 
\draw[->] [color={rgb, 255:red, 243; green, 93; blue, 45 }  ,draw opacity=1 ][line width=0.75]    (191.28,143.85) -- (234.95,143.52) ;

%Straight Lines [id:da5223923232586034] 
\draw[->] [line width=0.75]    (123.41,143.5) -- (123.41,183) ;

%Flowchart: Or [id:dp18526841832486174] 
\draw   (112.56,194.25) .. controls (112.56,188.04) and (117.46,183) .. (123.5,183) .. controls (129.54,183) and (134.44,188.04) .. (134.44,194.25) .. controls (134.44,200.46) and (129.54,205.5) .. (123.5,205.5) .. controls (117.46,205.5) and (112.56,200.46) .. (112.56,194.25) -- cycle ; \draw   (112.56,194.25) -- (134.44,194.25) ; \draw   (123.5,183) -- (123.5,205.5) ;
%Straight Lines [id:da5301862809832634] 
\draw [color={rgb, 255:red, 74; green, 144; blue, 226 }  ,draw opacity=1 ][line width=0.75]    (123.5,249) -- (123.5,205.5) ;
%Shape: Rectangle [id:dp3562392703508701] 
\draw  [color={rgb, 255:red, 243; green, 93; blue, 45 }  ,draw opacity=1 ][line width=0.75]  (161.18,172.81) -- (234,172.81) -- (234,216.76) -- (161.18,216.76) -- cycle ;
%Straight Lines [id:da7442498970518223] 
\draw[->] [color={rgb, 255:red, 243; green, 93; blue, 45 }  ,draw opacity=1 ][line width=0.75]    (211.39,143.68) -- (211.49,172.81) ;

%Straight Lines [id:da8707459934286401] 
\draw[->] [color={rgb, 255:red, 243; green, 93; blue, 45 }  ,draw opacity=1 ][line width=0.75]    (161,194) -- (134.44,194.23) ;

%Shape: Rectangle [id:dp73988367818608] 
\draw  [color={rgb, 255:red, 74; green, 144; blue, 226 }  ,draw opacity=1 ][line width=0.75]  (149.79,226.11) -- (196.9,226.11) -- (196.9,270.06) -- (149.79,270.06) -- cycle ;
%Straight Lines [id:da06959126871569565] 
\draw [->] [color={rgb, 255:red, 74; green, 144; blue, 226 }  ,draw opacity=1 ][line width=0.75]    (123.5,249) -- (149.79,249) ;

% Text Node
\draw (75.53,143.96) node  [font=\scriptsize] [align=left] {\begin{minipage}[lt]{40.43pt}\setlength\topsep{0pt}
\begin{center}
Resource\\de-mapping
\end{center}

\end{minipage}};
% Text Node
\draw (167.12,143.13) node  [font=\scriptsize,color={rgb, 255:red, 243; green, 93; blue, 45 } ] [align=left] {\begin{minipage}[lt]{25.74pt}\setlength\topsep{0pt}
\begin{center}
Demod
\end{center}

\end{minipage}};
% Text Node
\draw (212.26,132.38) node  [font=\scriptsize] [align=left] {LLR$\displaystyle _{\text{c}}$};
% Text Node
\draw (269.5,132.38) node  [font=\scriptsize] [align=left] {LLR$\displaystyle _{\text{c} ,k}$};
% Text Node
\draw (240,239.64) node  [font=\scriptsize] [align=left] {LLR$\displaystyle _{\text{p} ,k}$};
% Text Node
\draw (241.66,144.38) node  [font=\scriptsize,color={rgb, 255:red, 243; green, 93; blue, 45 }, rotate=-270] [align=left] {DEMUX};
% Text Node
\draw (296.5,193) node  [font=\scriptsize,rotate=-90] [align=left] {MUX};
% Text Node
\draw (363,192.55) node  [font=\scriptsize] [align=left] {\begin{minipage}[lt]{24.19pt}\setlength\topsep{0pt}
\begin{center}
FEC$\displaystyle ^{-1}$
\end{center}

\end{minipage}};
% Text Node
\draw (397,181.54) node  [font=\scriptsize] [align=left] {\begin{minipage}[lt]{13.73pt}\setlength\topsep{0pt}
\begin{center}
$\displaystyle \hat{W}_{k}$
\end{center}

\end{minipage}};
% Text Node
\draw (320.7,182) node  [font=\scriptsize] [align=left] {LLR$\displaystyle _{k}$};
% Text Node
\draw (114,128) node [anchor=north west][inner sep=0.75pt]  [font=\scriptsize] [align=left] {$\displaystyle y_{k}$};

\draw (105,170) node [anchor=north west][inner sep=0.75pt]  [font=\scriptsize] [align=left] {$+$};

\draw (130,177) node [anchor=north west][inner sep=0.75pt]  [font=\scriptsize] [align=left] {$-$};
% Text Node
\draw (196.99,195.13) node  [font=\scriptsize,color={rgb, 255:red, 243; green, 93; blue, 45 } ] [align=left] {\begin{minipage}[lt]{41.21pt}\setlength\topsep{0pt}
\begin{center}
Soft Symbol
\end{center}

\end{minipage}};
% Text Node
\draw (173.35,248.09) node  [font=\scriptsize,color={rgb, 255:red, 74; green, 144; blue, 226 }] [align=left] {\begin{minipage}[lt]{25.74pt}\setlength\topsep{0pt}
\begin{center}
Demod
\end{center}

\end{minipage}};

\end{tikzpicture}}
\caption{PHY receiver designs for CS-RSMA.}
\label{PHY RX}
\end{figure}

Fig. \ref{PHY RX} depicts two proposed receiver architectures for CS-RSMA, where ``FEC$^{-1}$" and ``Demod" represent channel decoder and demodulator (demapper) respectively. They are based on the joint de-mapping receiver and soft Symbol Level Interference Cancellation (SLIC) receiver for conventional RSMA proposed in \cite{Sibo_TCOM}. The two receiver architectures are explained as follows.
\begin{enumerate}
    \item Joint de-mapping: The receiver first jointly demodulates the common stream and the desired private stream from the receiver signal. Assuming a soft-output demodulator is in use, the demodulator produces Log-Likelihood Ratios (LLRs) for $X_\text{c}$ and $X_{\text{p},k}$, denoted as $\text{LLR}_\text{c}$ and $\text{LLR}_{\text{p},k}$. The receiver then de-multiplexes (extracts) the LLRs for $X_{\text{c},k}$ from $\text{LLR}_\text{c}$, denoted as $\text{LLR}_{\text{c},k}$. $\text{LLR}_{\text{c},k}$ and $\text{LLR}_{\text{p},k}$ are then multiplexed (combined) into LLRs of the desired unicast message, $X_k$, denoted as $\text{LLR}_k$. Finally, the receiver decodes $\text{LLR}_k$ into an estimate of the desired message $\Hat{W}_k$.
    \item Soft SLIC: Jointly demodulating $X_\text{c}$ and $X_{\text{p},k}$ may introduce a high computational burden, therefore, soft SLIC is proposed as a low-complexity alternative of joint demapping. With soft SLIC, the receiver first demodulates $s_\text{c}$ by treating all private streams as Gaussian noise and obtains $\text{LLR}_\text{c}$. Based on $\text{LLR}_\text{c}$, the receiver then computes the posterior probability distribution of the common stream symbols, denoted by $P(s_\text{c}|y)$. $P(s_\text{c}|y)$ is used to compute the soft symbol estimates, i.e., $\mathbb{E}[s_\text{c}|y]$, denoted by $\Tilde{s}$, and the variance of the residue noise, $\mathbb{E}[|s_\text{c} - \Tilde{s}|^2|y]$, denoted as $\sigma_{s_\text{c}}^2$. Next, the receiver applies soft cancellation of $s_\text{c}$ by subtracting $\mathbf{h}_k^H\mathbf{p}_\text{c}\Tilde{s}_\text{c}$ from $y_k$ and demodulates $s_{\text{p},k}$ from the remaining signal into $\text{LLR}_{\text{p},k}$. The remaining steps are the same as in the joint de-mapping implementation: the receiver de-multiplexes (extracts) $\text{LLR}_{\text{c},k}$ from $\text{LLR}_{\text{c}}$, multiplexes $\text{LLR}_{\text{c},k}$ and $\text{LLR}_{\text{p},k}$ into $\text{LLR}_k$, and decodes $\Hat{W}_k$ based on $\text{LLR}_k$.
\end{enumerate}

\subsection{Link-Level Simulation}
Link-level simulations were performed to evaluate the proposed CS-RSMA transceiver designs. In the following, we assume fast fading channels, where individual channel observations are generated using the one-ring channel model introduced in Section V, with $N_\text{T}=4$, $K=2$, and $\Delta=\pi/18$. For channel encoding and decoding, the LDPC code specified in 5G NR \cite{3gpp38212} with belief propagation decoding is chosen for its popularity and efficiency for long code lengths. All the BER results were obtained by averaging the results from 50000 blocks. The symbol block length is set to 512. For de-mapping, the max-log approximation was applied for its popularity in practice. To reduce simulation time, the low-complexity precoder design proposed in \cite{Sibo_TCOM} for RSMA without SIC is used for CS-RSMA. 

\begin{figure*}[!h]
\centering
    \includegraphics[width=14cm]{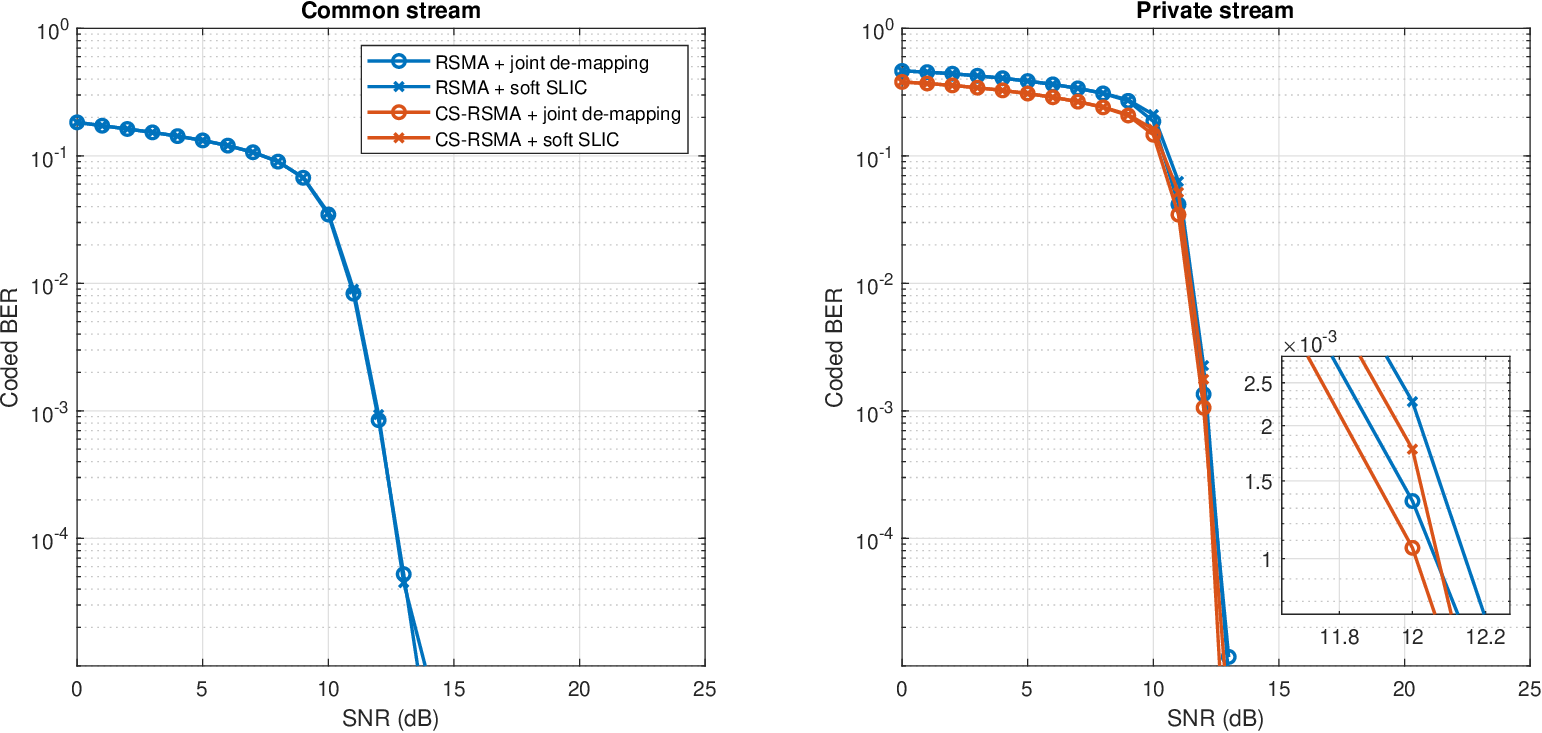}
    \caption{BER comparison between conventional RSMA and CS-RSMA with all streams using QPSK constellation. Code rate for RSMA: 0.85 for common streams, 0.325 for private streams. Code rate for CS-RSMA: 0.5 for both streams.}
    \label{BER_CS_RSMA}
\end{figure*}

Fig. \ref{BER_CS_RSMA} depicts BER curves of conventional RSMA and CS-RSMA with both joint de-mapping and soft SLIC receivers. It can be observed that, with the same receiver implementation, CS-RSMA performs slightly better than conventional RSMA. Taking $10^{-4}$ as the target BER, the SNR gap is around 0.05dB. The minor performance gap between CS-RSMA and conventional RSMA is consistent with the results from Section \ref{Sec: Opt results}. In addition, a soft SLIC receiver only leads to a minor performance loss compared to joint mapping, which is consistent with the observation in \cite{Sibo_TCOM}.

\subsection{Advantages of CS-RSMA}
Although the numerical results in Section V and this section show that the achievable rate and BER performance of CS-RSMA is not significantly different from conventional RSMA, the introduction of CS-RSMA is of significant importance for practical applications. We briefly discuss these benefits of implementing CS-RSMA compared to conventional RSMA in the following aspects.
\begin{enumerate}
    \item Reduced encoding/decoding complexity: With CS-RSMA, the transmitter encodes $K$ messages rather than $K+1$ messages as in conventional RSMA; the receiver decodes one codeword rather than two as in conventional RSMA. \rev{The $50\%$ reduction in the number of codewords for each user leads not only to reduced encoding/decoding complexity but also simplification on error detection (e.g., CRC check), interleaving, and scrambling.}
    \item Reduced signaling overhead: With CS-RSMA, each receiver only needs information on the encoding scheme for one codeword, rather than two (for common and private streams) as with conventional RSMA. This leads to fewer requirements on control signaling from CS-RSMA. \rev{The signaling overhead for modulation and coding is, in general, proportional to the number of active codewords. By reducing the total number of transmitted codewords from $K+1$ to $K$, CS-RSMA reduces the control information overhead for modulation and coding by approximately $1/(K+1)$.}
    \item Simplified retransmission design: With conventional RSMA, additional retransmission, e.g., Hybrid Automatic Repeat Request (HARQ), process needs to be implemented for the common stream. Due to the fact that the common stream is encoded with messages for potentially multiple users, designing the retransmission scheme for the common stream can be a complicated problem (some examples are given in \cite{Rafael}). With CS-RSMA, the retransmission mechanism only needs to be implemented for the $K$ encoded unicast messages, exactly the same as the widely used MU-MIMO. This means that the existence of CS-RSMA can be transparent to the retransmission process for conventional MU-MIMO. For example, to implement RSMA in 5G NR PDSCH channels, CS-RSMA is naturally compatible with the existing HARQ process, whereas conventional RSMA requires redesigning the HARQ process for the common stream.  
\end{enumerate}
Although this work only presents 1-layer CS-RSMA, i.e., with one common stream, the idea of codeword segmentation and creating common streams using part of segments can be trivially extended to multi-layer RSMA schemes, such as the hierarchical RSMA in \cite{Dai_HRS} and the general framework of multi-layer RSMA in \cite{Mao_EURASIP}, where more common streams need to be implemented. In these cases, the above three benefits will be significantly magnified as the number of common streams increases.

Overall, CS-RSMA brings significant convenience in terms of complexity in implementation and compatibility with existing systems, and performs slightly better than conventional RSMA without SIC.

\section{Conclusion}
We proposed a novel architecture for RSMA, namely CS-RSMA. To capture practical constraints on receiver complexity, we evaluated the performance of conventional RSMA and CS-RSMA using the GMI for FAGCI channels, which assumes that users implement suboptimal decoders that treat undesired signals as Gaussian random variables for lower complexity. We further formulated SR and MMF optimization problems and developed optimization algorithms for precoder design. Numerical results on SR and MMF evaluations reveal that CS-RSMA slightly outperforms conventional RSMA schemes in SR, while performing very similarly to them in MMF. Furthermore, PHY transceiver designs were proposed for CS-RSMA, and LLS were utilized to further evaluate CS-RSMA under the proposed transceiver implementations. Overall, we concluded that CS-RSMA performs very similarly to conventional RSMA, and occasionally outperforms it, both under information-theoretical metrics and LLS. Last but not least, we discussed the significant practical benefits of CS-RSMA on encoding/decoding complexity, control signalling overhead, and retransmission mechanism design in comparison to the conventional RSMA.

%---------------------------------------------------------------------%
%\newpage
\appendices

\section{Proof of Proposition \ref{c_opt_prop}}
$\mathcal{P}_{7}$ can be written as follows:
\begin{equation}
\begin{split}
    \mathcal{P}_{10}:\;\; \max_{\mathbf{c}}&\;\; \min_k\;\; c_i I_{\text{c},k}' + I_{\text{p},k}\\
    \text{s.t.}&\;\;\;\; \mathbf{c} \succcurlyeq \mathbf{0},\\
    &\;\;\;\; \mathbf{1}^T \mathbf{c} = 1.\\
\end{split}
\end{equation}
For conventional RSMA (either with or without SIC), $I_{\text{c},k}' =\min_k I_{\text{c},k}$, $\forall k\in\mathcal{K}$. For CS-RSMA, $I_{\text{c},k}' = I_{\text{c},k}$, , $\forall k\in\mathcal{K}$.

By introducing a slack variable, $\xi$, $\mathcal{P}_{10}$ can be converted into
\begin{equation}
\begin{split}
    \mathcal{P}_{11}:\;\; \max_{\mathbf{c}}&\;\;\;\;\;\; \xi\\
    \text{s.t.}&\;\;\;\; \mathbf{c} \succcurlyeq \mathbf{0},\\
    &\;\;\;\; \mathbf{c}^T \mathbf{1} = 1,\\
    &\;\;\;\; c_i I_{\text{c},k}' + I_{\text{p},k} \geq \xi,\; \forall k.
\end{split}
\end{equation}

WLOG, we assume $I_{\text{p},1}<I_{\text{p},2}<...<I_{\text{p},K}$. Assume that the solution of $\mathcal{P}_{11}$ leads to $c_i>0$ for $k=1,\;...,\;k'$. It is easy to see that $c_i I'_{\text{c},k} + I_{\text{p},k} = \xi$ for $k=1,...,k'$. Equivalently,
\begin{equation}\label{linear equation}
    \mathbf{Ac}' + \mathbf{b} = \xi\mathbf{1},
\end{equation}
where $\mathbf{A} = \text{diag}\{[I'_{\text{c},1},\;...,\;I'_{\text{c},k'}]\}$, $\mathbf{b}=[I_{\text{p},1},\;...,\;I_{\text{p},k'}]^T$, $\mathbf{c}'=[c_1,\;...,\;c_{k'}]^T$, and $\mathbf{1}$ is a column vector of length $k'$ whose elements are all ones. The global optimal solution of $\mathbf{c}'$ and $\xi$, $\xi^\star$ and $\mathbf{c}'^\star$, can then be obtained by
\begin{equation}\label{c_opt}
    \mathbf{c}'^\star = \xi\mathbf{A}^{-1}\mathbf{1} - \mathbf{A}^{-1}\mathbf{b}
\end{equation}
and
\begin{equation}\label{t_opt}
    \xi^\star = \frac{1+\mathbf{1}^T\mathbf{A}^{-1}\mathbf{b}}{\mathbf{1}^T\mathbf{A}^{-1}\mathbf{1}}.
\end{equation}

Note that (\ref{c_opt}) is directly from (\ref{linear equation}), and (\ref{t_opt}) is from (\ref{c_opt}) combined the condition of $\mathbf{1}^T \mathbf{c} = 1$.

Although $k'$, i.e., the number of users who are strictly served by the common stream, is still unknown, we can test it using (\ref{c_opt}) and (\ref{t_opt}). This is because allocating the common stream resource to more users than $k'$ will lead to $\mathbf{c}'$ containing negative entries, which is infeasible. Hence, by iteratively guessing the number of users to be served by the common stream, Alg. \ref{alg: c_opt} is obtained.

\section{\rev{Proof of the Achievability of \eqref{equ: rate CS-RSMA}}}\label{appdx: achievability proof}
\rev{
We consider a CS-RSMA decoder at user-$k$ that takes $\mathbf{y}_k$ as the input and outputs an estimate of the message, which is written in the following form
\begin{equation}
    \Hat{m} = \underset{m=1,...,M}{\arg\;\max} q(\mathbf{s}_{\text{c},k}^m,\mathbf{s}_{\text{p},k}^m,\mathbf{y}_k),
\end{equation}
where $[\mathbf{s}_{\text{c},k}^m;\;\mathbf{s}_{\text{p},k}^m] = \mathbf{s}_{k}^{\text{cw},m}$ is the codeword for user-$k$ corresponding to message $m$. We assume without loss of generality that $\mathbf{s}_{\text{c},k}$ occupies $\mathbf{s}_{\text{c}}$ at $t=1,...,n_k-n$. \eqref{equ: Rx signal} can be then rewritten as
\begin{equation}\label{equ: achievability proof channel 1}
    y_{t,k} = \mathbf{h}_k^H\mathbf{p}_\text{c} s_{t,\text{c},k} +  \mathbf{h}_k^H\mathbf{p}_{\text{p},k} s_{t,\text{p},k} + \Tilde{z}_{t,k}, \text{\;\;for\;\;} t=1,...,n_k-n
\end{equation}
and
\begin{multline}\label{equ: achievability proof channel 2}
    y_{t,k} = \mathbf{h}_k^H\mathbf{p}_{\text{p},k} s_{t,\text{p},k} + \mathbf{h}_k^H\mathbf{p}_\text{c} s_{t,\text{c},k'} + \Tilde{z}_{t,k},\\ \text{\;\;for\;\;} t=n_k-n +1 ,...,n \text{\;\;and\;} k'\in\mathcal{K}/k,
\end{multline}
where $s_{t,\text{p},k}$ is the private stream symbol intended for user-$k$ at time instance $t$, $s_{t,\text{c},k}$ is the common stream symbol at time instance $t$ and is from $\mathbf{s}_{\text{c},k}$, and $\Tilde{z}_{t,k}$ contains interference from the private streams intended for other users and the AWGN noise, i.e.,
\begin{equation}
    \Tilde{z}_{t,k} = \sum_{k^\circ\in\mathcal{K}/k}\mathbf{h}_{k}^H\mathbf{p}_{\text{p},k^\circ} s_{t,\text{p},k^\circ} + z_{t,k}.
\end{equation}
Since the channel is memoryless across different time indices, we consider a decoding metric that can be decomposed as follows.
\begin{equation}
\begin{split}
    \Hat{m} &= \underset{m=1,...,M}{\arg\;\max} q(\mathbf{s}_{\text{c},k}^m,\mathbf{s}_{\text{p},k}^m,\mathbf{y}_k)\\
    &= \underset{m=1,...,M}{\arg\;\max} \prod_{t=1}^{n_k- n}q_a(s_{t,\text{c},k}^m,s_{t,\text{p},k}^m,y_{t,k}) \prod_{t=n_k- n+1}^{n_k}q_b(s_{t,\text{p},k}^m,y_{t,k})^{\rho_b}\\
\end{split}
\end{equation}
We further consider suboptimal decoding metrics that are separable in $s_{t,\text{c},k}$ and $s_{t,\text{p},k}$, i.e., 
\begin{equation}
    q_a(s_{t,\text{c},k},s_{t,\text{p},k},y_{t,k}) = q_{a,\text{c}}(s_{t,\text{c},k},y_{t,k})^{\rho_{a,\text{c}}}q_{a,\text{p}}(s_{t,\text{p},k},y_{t,k})^{\rho_{a,\text{p}}}.
\end{equation}
$\rho_{a,\text{c}}$, $\rho_{a,\text{p}}$ and $\rho_{b}$ are positive parameters. We evaluate the achievable rate using the GMI. With capital letters representing the random variables and the time indices omitted when unnecessary, the GMI for user-$k$ can be derived as in \eqref{equ: achievability proof GMI}.}

\begin{figure*}
\rev{
    \begin{equation}\label{equ: achievability proof GMI}
        \begin{split}
            I_\text{GMI} 
        =& \underset{s>0,\rho_{a,\text{c}}>0,\rho_{a,\text{p}}>0,\rho_{b}>0}{\sup}\;\;\; \frac{1}{n}\mathbb{E} \left[ \log\; \frac{q(\mathbf{S}_{\text{c},k},\mathbf{S}_{\text{p},k},\mathbf{Y}_k)^s}{\mathbb{E} \left[ q(\Bar{\mathbf{S}}_{\text{c},k},\Bar{\mathbf{S}}_{\text{p},k},\mathbf{Y}_k)^s \mid \mathbf{Y}_k)\right ]} \right]\\
        =& \underset{s>0,\rho_{a,\text{c}}>0,\rho_{a,\text{p}}>0,\rho_{b}>0}{\sup}\;\;\; \frac{1}{n}\mathbb{E} \left[ \log\; \frac{\prod_{t=1}^{n_k- n}q_{a,\text{c}}(S_{t,\text{c},k},Y_{t,k})^{\rho_{a,\text{c}}s}q_{a,\text{p}}(S_{t,\text{p},k},Y_{t,k})^{\rho_{a,\text{p}}s} \prod_{t=n_k- n+1}^{n_k}q_b(S_{t,\text{p},k},Y_{t,k})^{\rho_{b}s}}{\mathbb{E} \left[ \prod_{t=1}^{n_k- n}q_{a,\text{c}}(\Bar{S}_{t,\text{c},k},Y_{t,k})^{\rho_{a,\text{c}}s}q_{a,\text{p}}(\Bar{S}_{t,\text{p},k},Y_{t,k})^{\rho_{a,\text{p}}s} \prod_{t=n_k- n+1}^{n_k}q_b(\Bar{S}_{t,\text{p},k},Y_{t,k})^{\rho_{b}s} \mid \mathbf{Y}_k)\right ]} \right]\\
        =& \underset{s_1>0,s_2>0,s_3>0}{\sup}\\
        &\frac{1}{n}\mathbb{E} \left[ \log\; \frac{\prod_{t=1}^{n_k- n}q_{a,\text{c}}(S_{t,\text{c},k},Y_{t,k})^{s_1}q_{a,\text{p}}(S_{t,\text{p},k},Y_{t,k})^{s_2} \prod_{t=n_k- n+1}^{n_k}q_b(S_{t,\text{p},k},Y_{t,k})^{s_3}}{\mathbb{E} \left[ \prod_{t=1}^{n_k- n}q_{a,\text{c}}(\Bar{S}_{t,\text{c},k},Y_{t,k})^{s_1}\mid \mathbf{Y}_k)\right ]\mathbb{E} \left[ \prod_{t=1}^{n_k- n}q_{a,\text{p}}(\Bar{S}_{t,\text{p},k},Y_{t,k})^{s_2} \mid \mathbf{Y}_k)\right ] \mathbb{E} \left[\prod_{t=n_k- n+1}^{n_k}q_b(\Bar{S}_{t,\text{p},k},Y_{t,k})^{s_3} \mid \mathbf{Y}_k)\right ]} \right]\\
        =& \underset{s>0}{\sup}\;\;\;
        \frac{1}{n}\mathbb{E} \left[ \log\; \frac{\prod_{t=1}^{n_k- n}q_{a,\text{c}}(S_{t,\text{c},k},Y_{t,k})^{s}}{\mathbb{E} \left[ \prod_{t=1}^{n_k- n}q_{a,\text{c}}(\Bar{S}_{t,\text{c},k},Y_{t,k})^{s}\mid \mathbf{Y}_k)\right ]} \right]
        + \underset{s>0}{\sup}\;\;\; \frac{1}{n}\mathbb{E} \left[ \log\; \frac{\prod_{t=1}^{n_k- n}q_{a,\text{p}}(S_{t,\text{p},k},Y_{t,k})^{s}}{\mathbb{E} \left[ \prod_{t=1}^{n_k- n}q_{a,\text{p}}(\Bar{S}_{t,\text{p},k},Y_{t,k})^{s} \mid \mathbf{Y}_k)\right ] } \right]\\
        &+\underset{s>0}{\sup}\;\;\; \frac{1}{n}\mathbb{E} \left[ \log\; \frac{\prod_{t=n_k- n+1}^{n_k}q_b(S_{t,\text{p},k},Y_{t,k})^{s}}{\mathbb{E} \left[\prod_{t=n_k- n+1}^{n_k}q_b(\Bar{S}_{t,\text{p},k},Y_{t,k})^{s} \mid \mathbf{Y}_k)\right ]} \right]\\
        =& \underset{s>0}{\sup}\;\;\;
        \frac{n_k-n}{n}\mathbb{E} \left[ \log\; \frac{q_{a,\text{c}}(S_{t,\text{c},k},Y_{t,k})^{s}}{\mathbb{E} \left[ q_{a,\text{c}}(\Bar{S}_{t,\text{c},k},Y_{t,k})^{s}\mid Y_k)\right ]} \right]
        + \underset{s>0}{\sup}\;\;\; \frac{n_k-n}{n}\mathbb{E} \left[ \log\; \frac{q_{a,\text{p}}(S_{t,\text{p},k},Y_{t,k})^{s}}{\mathbb{E} \left[q_{a,\text{p}}(\Bar{S}_{t,\text{p},k},Y_{t,k})^{s} \mid Y_k)\right ] } \right]\\
        &+\underset{s>0}{\sup}\;\;\; \frac{2n-n_k}{n}\mathbb{E} \left[ \log\; \frac{q_b(S_{t,\text{p},k},Y_{t,k})^{s}}{\mathbb{E} \left[q_b(\Bar{S}_{t,\text{p},k},Y_{t,k})^{s} \mid Y_k)\right ]} \right]\\
        =& \underset{s>0}{\sup}\;\;\;
        c_n \mathbb{E} \left[ \log\; \frac{q_{a,\text{c}}(S_{t,\text{c},k},Y_{t,k})^{s}}{\mathbb{E} \left[ q_{a,\text{c}}(\Bar{S}_{t,\text{c},k},Y_{t,k})^{s}\mid Y_k)\right ]} \right]
        + \underset{s>0}{\sup}\;\;\; c_n \mathbb{E} \left[ \log\; \frac{q_{a,\text{p}}(S_{t,\text{p},k},Y_{t,k})^{s}}{\mathbb{E} \left[q_{a,\text{p}}(\Bar{S}_{t,\text{p},k},Y_{t,k})^{s} \mid Y_k)\right ] } \right]\\
        &+\underset{s>0}{\sup}\;\;\; (1 - c_n) \mathbb{E} \left[ \log\; \frac{q_b(S_{t,\text{p},k},Y_{t,k})^{s}}{\mathbb{E} \left[q_b(\Bar{S}_{t,\text{p},k},Y_{t,k})^{s} \mid Y_k)\right ]} \right]\\
\end{split}
\end{equation}}
\end{figure*}
\rev{
Under the channel model in (\ref{equ: achievability proof channel 1}) and (\ref{equ: achievability proof channel 2}) and the suboptimal decoding setup given in Section \ref{subsec: decoding strategy and GMI}, i.e., treating undesired private streams as Gaussian random variables, we have 
\begin{equation}\label{equ: achi proof common rate}
    \underset{s>0}{\sup}\;\;\; \mathbb{E} \left[ \log\; \frac{q_{a,\text{c}}(S_{t,\text{c},k},Y_{t,k})^{s}}{\mathbb{E} \left[ q_{a,\text{c}}(\Bar{S}_{t,\text{c},k},Y_{t,k})^{s}\mid Y_k)\right ]} \right] = I_{\text{c},k}    
\end{equation}
and
\begin{equation}\label{equ: achi proof private rate}
\begin{split}
    &\underset{s>0}{\sup}\;\;\; \mathbb{E} \left[ \log\; \frac{q_{a,\text{p}}(S_{t,\text{p},k},Y_{t,k})^{s}}{\mathbb{E} \left[q_{a,\text{p}}(\Bar{S}_{t,\text{p},k},Y_{t,k})^{s} \mid Y_k)\right ] } \right]\\
    =& \underset{s>0}{\sup}\;\;\; \mathbb{E} \left[ \log\; \frac{q_b(S_{t,\text{p},k},Y_{t,k})^{s}}{\mathbb{E} \left[q_b(\Bar{S}_{t,\text{p},k},Y_{t,k})^{s} \mid Y_k)\right ]} \right]\\
    =& I_{\text{p},k}^{\text{\rev{SIC-free}}},
\end{split}
\end{equation}
(which was proved in \cite{Sibo_FAGCI}). Substituting \eqref{equ: achi proof common rate} and \eqref{equ: achi proof private rate} into \eqref{equ: achievability proof GMI}, we obtain \eqref{equ: rate CS-RSMA}.}

\section*{Acknowledgment}
\rev{Sibo Zhang and Bruno Clerckx would like to express their sincere gratitude to Dr. Hamdi Joudeh, with the Department of Electrical Engineering at Eindhoven University of Technology, for his insightful comments on this work and for the valuable discussions regarding the achievability proof in Appendix B.}

\bibliographystyle{IEEEtran}
\bibliography{references}

\end{document}